%% file: main.tex
\documentclass[sigconf]{acmart}

\AtBeginDocument{%
  \providecommand\BibTeX{{%
    Bib\TeX}}}

\copyrightyear{2025}
\acmYear{2025}
\setcopyright{acmlicensed}\acmConference[MICRO '25]{58th IEEE/ACM International Symposium on Microarchitecture}{October 18--22, 2025}{Seoul, Republic of Korea}
\acmBooktitle{58th IEEE/ACM International Symposium on Microarchitecture (MICRO '25), October 18--22, 2025, Seoul, Republic of Korea}
\acmDOI{10.1145/3725843.3756025}
\acmISBN{979-8-4007-1573-0/2025/10}

\usepackage{acmart-taps}

\usepackage{amssymb,amsfonts}
\usepackage{textcomp}
\usepackage{multirow}
\usepackage{makecell}
\usepackage{algorithm}
\usepackage{pifont}
\usepackage{xspace}

\newcommand{\sysname}{\ensuremath{\mathsf{Ironman}}\xspace}
\newcommand{\xor}{\ensuremath{\oplus}}

\def\BibTeX{{\rm B\kern-.05em{\sc i\kern-.025em b}\kern-.08em
    T\kern-.1667em\lower.7ex\hbox{E}\kern-.125emX}}

\begin{document}

\title{Ironman: Accelerating Oblivious Transfer Extension for Privacy-Preserving AI with Near-Memory Processing}





\author{Chenqi Lin}
\affiliation{%
  \institution{Peking University}
  \city{Beijing}
  \country{China}}
\email{linchenqi@stu.pku.edu.cn}

\author{Kang Yang}
\affiliation{%
  \institution{State Key Laboratory of Cryptology}
  \city{Beijing}
  \country{China}}
\email{yangk@sklc.org}

\author{Tianshi Xu}
\affiliation{%
  \institution{Peking University}
  \city{Beijing}
  \country{China}}
\email{tianshixu@stu.pku.edu.cn}

\author{Ling Liang}
\affiliation{%
  \institution{Peking University}
  \city{Beijing}
  \country{China}}
\email{lingliang@pku.edu.cn}

\author{Yufei Wang}
\affiliation{%
    \institution{DAMO Academy, Alibaba Group}
    \city{Hangzhou}
    \country{China}}
\affiliation{%
    \institution{Hupan Lab}
    \city{Hangzhou}
    \country{China}}
\email{wangyufei.wyf@alibaba-inc.com}

\author{Zhaohui Chen}
\affiliation{%
    \institution{DAMO Academy, Alibaba Group}
    \city{Hangzhou}
    \country{China}}
\affiliation{%
    \institution{Hupan Lab}
    \city{Hangzhou}
    \country{China}}
\email{chenzhaohui.czh@alibaba-inc.com}

\author{Runsheng Wang}
\affiliation{%
  \institution{Peking University}
  \city{Beijing}
  \country{China}}
\email{wrs@pku.edu.cn}

\author{Mingyu Gao}
\affiliation{%
  \institution{Tsinghua University}
  \city{Beijing}
  \country{China}}
\email{gaomy@tsinghua.edu.cn}

\author{Meng Li}
\authornote{Corresponding author.}
\affiliation{%
  \institution{Peking University}
  \city{Beijing}
  \country{China}}
\email{meng.li@pku.edu.cn}

\ccsdesc[500]{Computer systems organization~Special purpose systems}
\ccsdesc[500]{Security and privacy~Cryptography}

\keywords{Oblivious Transfer, Learning Parity with Noise, Near Memory Processing}

\renewcommand{\shortauthors}{Meng Li, et al.}

\input{docs/0-abstract}

\maketitle

\input{docs/1-introduction_v1}

\input{docs/2-background}

\input{docs/3-motivation_v2}

\input{new_method_2/spcot}
\input{new_method_2/lpn}

\input{docs/6-setup}

\input{docs/7-experiment}

\input{docs/conclusion}



\bibliographystyle{ACM-Reference-Format}
\bibliography{abbrev3,crypto,reference}

\end{document}

%% file: docs/0-abstract.tex
\begin{abstract}



With the wide application of machine learning (ML), privacy concerns arise with user data as they may contain sensitive information. Privacy-preserving ML (PPML) based on cryptographic primitives has emerged as a promising solution in which an ML model is directly computed on the encrypted data to provide a formal privacy guarantee. However, PPML frameworks heavily rely on the oblivious transfer (OT) primitive to compute nonlinear functions. OT mainly involves the computation of single-point correlated OT (SPCOT) and learning parity with noise (LPN) operations. As OT is still computed extensively on general-purpose CPUs, it becomes the latency bottleneck of modern PPML frameworks.


In this paper, we propose a novel OT accelerator, dubbed~\sysname, to significantly increase the efficiency of OT and the overall PPML framework. We observe that SPCOT is computation-bounded, and thus propose a hardware-friendly SPCOT algorithm with a customized accelerator to improve SPCOT computation throughput. In contrast, LPN is memory-bandwidth-bounded due to irregular memory access patterns. Hence, we further leverage the near-memory processing (NMP) architecture equipped with memory-side cache and index sorting to improve effective memory bandwidth. With extensive experiments, we demonstrate \sysname~achieves a $39.2$–$237.4\times$ improvement in OT throughput across different NMP configurations compared to the full-thread CPU implementation. For different PPML frameworks, \sysname demonstrates a $2.1$–$3.4 \times$ reduction in end-to-end latency for both CNN and Transformer models.

\end{abstract}


%% file: docs/1-introduction_v1.tex
\section{Introduction}
\label{sec:intro}
%



Machine learning has demonstrated state-of-the-art (SOTA) performance across a wide range of tasks and is increasingly being adopted in sensitive and privacy-critical applications. However, to leverage machine learning in cloud environments, users are often required to disclose their inputs directly to the service provider, which may include sensitive information such as health records and location data \cite{sweeney2002k}. As a result, privacy has become a significant concern. To address these concerns, Privacy-Preserving Machine Learning (PPML) techniques, including Homomorphic Encryption (HE) \cite{gentry2009fully} and Multi-Party Computation (MPC) \cite{demmler2015aby, SP:BHSSY23}, have recently emerged as promising solutions.



The hybrid HE/MPC frameworks~\cite{rathee2020cryptflow2,huang2022cheetah,hao2022iron,pang2024bolt,lu2024bumblebee} are widely adopted in PPML, with Oblivious Transfer (OT) serving as a fundamental component for the evaluation of nonlinear activation functions. In contrast to Fully Homomorphic Encryption (FHE)-based frameworks—another prevalent approach in PPML, the OT-based protocol enables more accurate computation of complex nonlinear layers (e.g., Softmax, GeLU) in Transformer models.  Consequently, the OT-based protocol does not degrade inference accuracy and eliminates the need for extensive fine-tuning.
However, hybrid HE/MPC frameworks require the generation of a large number of OT correlations for private inference, which incurs substantial communication costs. To address this issue, the PCG-style OT Extension (OTE) has been introduced~\cite{CCS:BCGI18,C:BCGIKS19,CCS:BCGIKR19,CCS:YWLZW20,C:CouRinRag21,C:BCGIKR22,EC:GYWZXZ23,C:RagRinTan23}. This method enables two parties to extend a small number of PKC-based OT correlations into a large number of OT correlations, with sub-linear communication complexity. Although this significantly alleviates the pressure on the communication bandwidth, it comes at the cost of increased computational overhead.


\begin{figure}[!tb]
    \centering
    \includegraphics[width=1.0\linewidth]{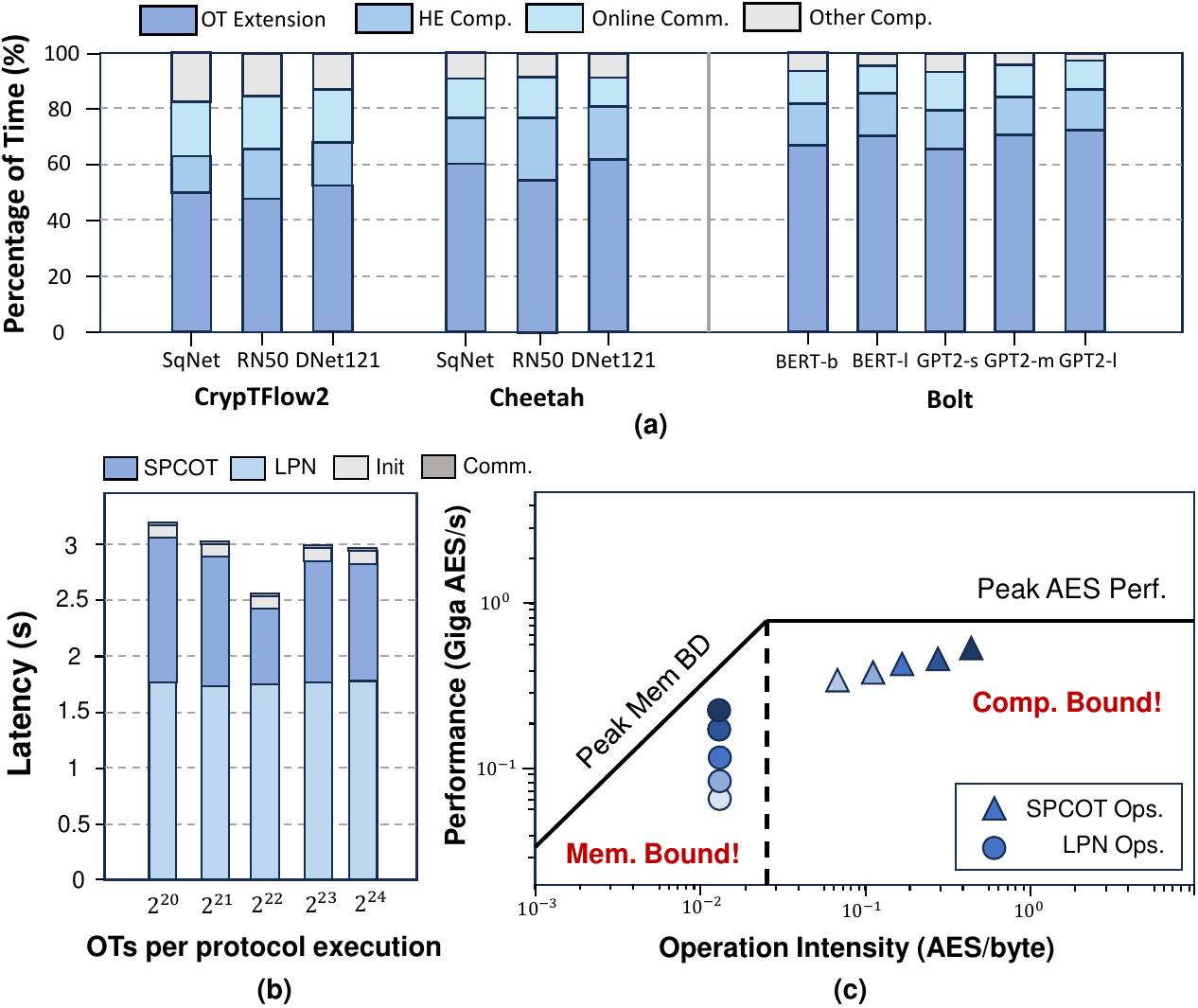}
    \caption{(a) Percentage of execution time for each model across distinct PPML frameworks. (b) Latency comparison across different parameters, with the number of OTs ranging from $2^{20}$ to $2^{24}$ (about $2^{25}$ OTs required by the first layer in secure ResNet18 inference). The parameters are detailed in Table~\ref{tab:parameter_sets}.
    (c) Roofline model of multi-threaded SPCOT and LPN encoding. The number of OTs output by each protocol execution is increased from $2^{20}$ to $2^{24}$, with darker colors representing a larger number of OTs.}
    \label{fig:intro}
    \Description{Introduction}
\end{figure}

Here, we conduct an experiment to identify the bottleneck of SOTA hybrid HE/MPC frameworks for both the CNN and Transformer models. Figure~\ref{fig:intro}(a) illustrates the percentage of execution time across different components. We observe that, with GPU acceleration and optimized protocols, the latency bottleneck is not associated with the evaluation of linear layers, but rather with the latency of the PCG-style OTE, which accounts for 51\% to 69\% of the total execution time across all models and frameworks. This is due to the lack of existing work that addresses the mitigation of PCG-style OTE latency from a hardware perspective.

A PCG-style OTE protocol mainly includes two components: the Single-Point Correlated OT (SPCOT) sub-protocol and Learning Parity with Noise (LPN) encoding. In particular, SPCOT makes two parties distributedly compute hundreds of Goldreich-Goldwasser-Micali (GGM) trees~\cite{GGM86}, where a GGM tree uses a Pseudo-Random Generator (PRG) to expand a parent node into two child nodes. All known PCG-style OTE implementations adopt AES to instantiate PRG, due to the AES-NI acceleration on CPU.
The LPN encoding involves performing numerous random accesses to a vector with millions of components.
We profile the SOTA PCG-style OTE protocol~\cite{CCS:YWLZW20} and make the following observation based on Figure~\ref{fig:intro}(b): the protocol latency is primarily incurred by the computation of SPCOT and LPN, which are the focus of our OTE accelerator.
The main computational operations involved in both SPCOT and LPN are to perform a mass of AESs: SPCOT uses it to generate a GGM tree, while LPN uses it to generate indices of random access. Therefore, we use the number of AESs per second to measure the performance of SPCOT and LPN encoding in Figure~\ref{fig:intro} (c). As shown in Figure~\ref{fig:intro} (c), SPCOT is computation-bounded, while the LPN encoding is memory-bandwidth-bounded. Therefore, we need to use distinct strategies to accelerate SPCOT and LPN. 

In most OT-based MPC applications, two parties execute two OTE protocols in parallel when switching roles of the sender and receiver in the second protocol execution. The parallel OTE execution allows us to reduce the protocol latency. 
As the sender and receiver perform different operations in PCG-style OTE, we need to design a unified architecture for the OTE accelerator, which supports the same party to switch the roles of sender and receiver.


In this paper, we propose \sysname, a framework featured algorithm and hardware co-optimization to accelerate the PCG-style OTE. We introduce Hardware-aware m-ary GGM Tree Expansion to reduce computation overhead of SPCOT, which is applicable to hardware environments both with and without AES-NI instructions. Alongside the optimizations, we present a NMP architecture designed to support both SPCOT and LPN operations. \sysname~employs rank-level parallelism and an index sorting algorithm assisted with a memory-side cache to minimize the random access of LPN. Finally, given that sender and receiver have a different algorithm procedure, we design a unified unit for role switching scenario. 
The contributions of this work are summarized as follows:
\begin{itemize}
    \item To the best of our knowledge, we propose the first customized accelerator for the PCG-style OTE, dubbed \sysname, which features near-DRAM processing for SPCOT and LPN with a unified architecture to support both sender and receiver in OTE.
    \item For SPCOT, \sysname~leverages hardware-friendly PRG and compute-efficient m-ary tree expansion algorithm to mitigate the computation bottleneck.
    \item \sysname~features rank-level parallelism as well as an index sorting algorithm assisted with a memory-side cache, leading to high effective bandwidth for LPN acceleration.
    \item \sysname achieves $39.2\times - 237.4\times$ speedup over the current OTE implementation on CPUs. When applied to privacy-preserving machine learning, \sysname~enables $2.1\times - 3.4\times$ across different hybrid HE/MPC frameworks for both CNN and Transformer models.
\end{itemize}

%% file: docs/2-background.tex
\section{Background}

%

\begin{figure}[!tb]
    \centering
    \includegraphics[width=0.9\linewidth]{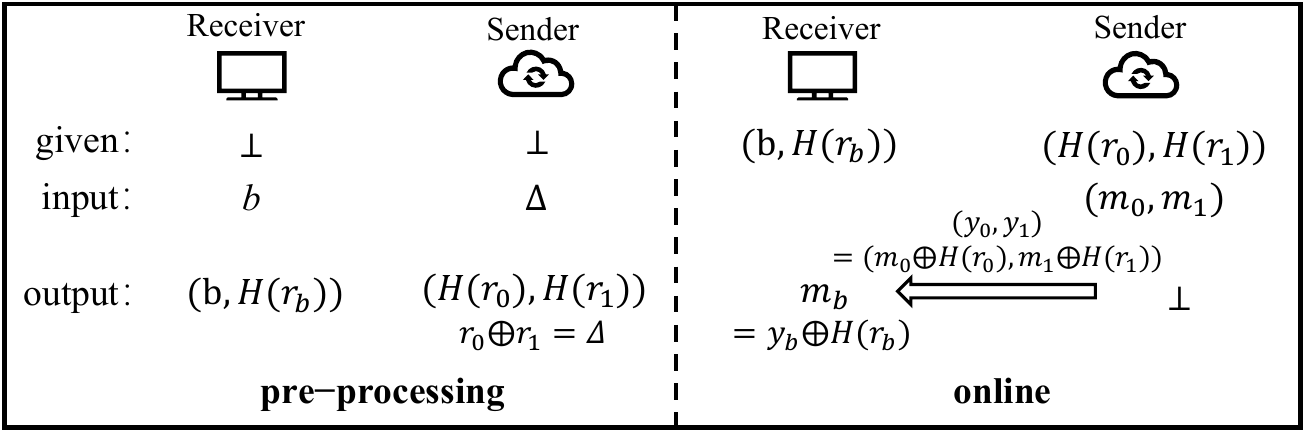}
    \caption{The Oblivious Transfer framework with COT correlations. Here, $\bot$ denotes NaN.}
    \label{fig:ot_definition}
    \Description{OT Definition}
\end{figure}

\begin{figure*}[!tb]
    \centering
    \includegraphics[width=0.9\linewidth]{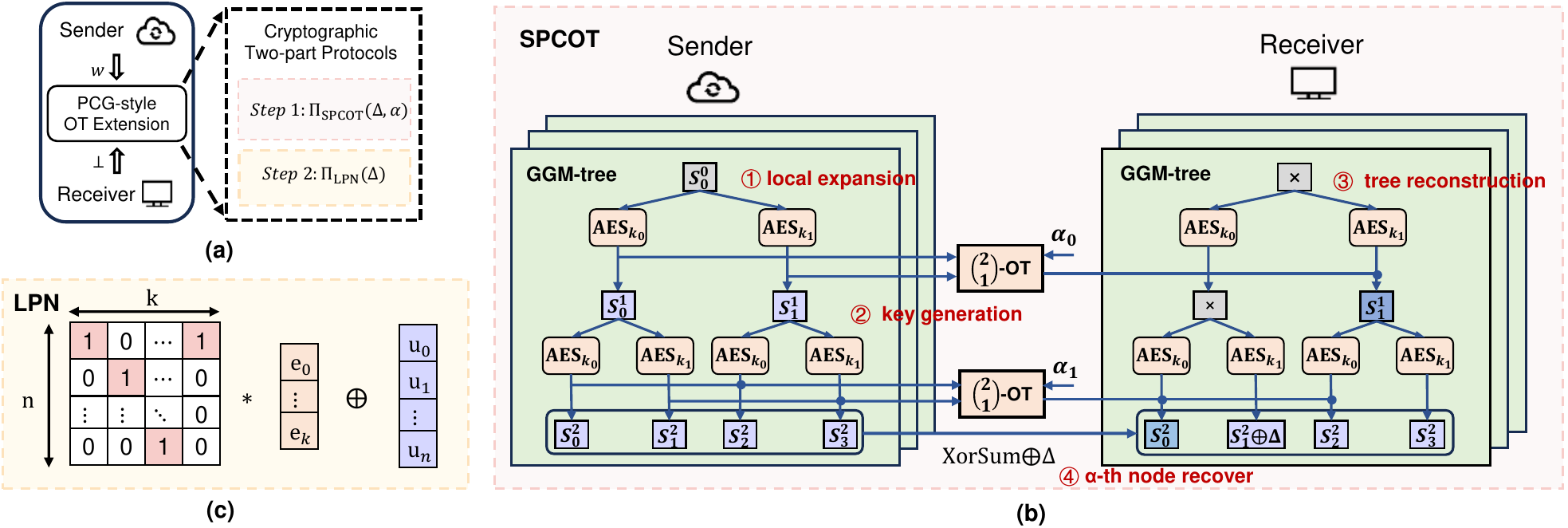}
    \caption{(a) Illustration of the PCG-style OTE algorithm. (b) A simplified example of $4$-node GGM tree generation in SPCOT operation. (c) LPN operation
    }
    \label{fig:vole_process}
    \Description{PCG-style OT Extension Protocol}
\end{figure*}

\subsection{Oblivious Transfer}
\label{subsec:ot}
%
Oblivious Transfer (OT) is a fundamental cryptographic primitive and has been used to construct a large number of protocols, e.g., Secure Multi-Party Computation (MPC)~\cite{FOCS:Yao86,STOC:GolMicWig87,C:NNOB12,AC:FKOS15,AC:HazSchSor17,C:KRRW18,AC:HeaKolPec20,USENIX:PSSY21,EC:CWYY23,demmler2015aby,SP:BHSSY23,CCS:HLWHWC24}, Zero-Knowledge Proofs (ZKPs)~\cite{SP:WYKW21,DIO21,C:BMRS21,CCS:YSWW21,CCS:FKLOWW21,CCS:BBMRS21,USENIX:WYXKW21,CCS:DILO22,CCS:YHHKV23,DCC:BDSW23,HHKVY24}, Private Set Intersection (PSI)~\cite{USENIX:PinSchZoh14,EC:RinRos17,EC:PSWW18,C:PRTY19,EC:PSTY19,CM20,EC:PRTY20,EC:RinSch21,C:GPRTY21,CCS:RagRin22}, particularly in privacy preserving machine learning inference and training~\cite{SP:NWIJBT13,SP:MohZha17,USENIX:JuvVaiCha18,CCS:SGRP19,CCS:ASKG19,USENIX:RSCLLK19,CCS:RRKCGR20,USENIX:MLSZP20,SP:RRGGSC21,USENIX:HLHD22,USENIX:CGOS22,USENIX:LHZWH23,HLCXXZ22,EPRINT:HLLLLH23,SP:PZMZS24,LHG+25}
Informally, OT allows a sender to input two messages $m_0,m_1$ and a receiver to input a bit $b$, and then lets the receiver obtain the message $m_b$. For security, $b$ is kept secret against the sender, and the receiver learns nothing about $m_{1-b}$. 

This requirement of OT can be met by employing Correlated OT (COT) correlations. Figure~\ref{fig:ot_definition} presents a definition of COT correlations and illustrates their application in implementing oblivious transfer. 
In this framework, an COT correlation is defined such that the sender obtains two random strings $r_0, r_1$ satisfying $r_1\oplus r_0=\Delta$,where $\Delta$ is fixed across multiple correlations. Simultaneously, the receiver obtains a random bit $b$ and the corresponding string $r_b=r_0\oplus b \Delta$. After generating COT correlations, they are converted into standard OT correlations using a Correlation Robust Hash Function (CRHF)~\cite{C:IKNP03}, denoted as $H$. In the online phase, the sender uses $H(r_0)$ and $H(r_1)$ to mask the message $m_0$ and $m_1$ respectively before transmitting them to the receiver, who then employs the pre-computed $H(r_b)$ to decrypt the selected message. 

\subsection{OT in PPML}
\label{subsec:OT_in_PPML}

OT is primarily used for non-linear function evaluation in PPML~\cite{rathee2020cryptflow2, rathee2021sirnn, huang2022cheetah, hao2022iron, pang2024bolt, lu2024bumblebee, srinivasan2019delphi}. The non-linear protocol also consists of two sequential steps:
\textbf{\ding{172} Pre-processing phase}, in which COT correlations are generated by OT Extension (OTE) and converted to OT correlations, and
\textbf{\ding{173} Online OT protocol}, evaluates non-linear functions by relying on both pre-generated OT and the secret sharing scheme.
Specifically, online OT protocols take secret shares of the input and return secret shares of the result after applying a nonlinear function, such as comparison, truncation, or table lookup. These building blocks are further used to implement common activation functions like ReLU and GELU.

The detailed design of how the OT protocol is specifically deployed for non-linear function evaluation has been extensively studied in prior works, which involve complex protocols. For a more detailed explanation, we refer readers to works such as~\cite{rathee2020cryptflow2, rathee2021sirnn}, which provide in-depth descriptions of how OT protocols are integrated with secret-shared inputs to securely evaluate non-linear functions.
Since OTE is the dominant cost in the preprocessing phase and the online OT protocol is already efficient, \sysname focuses on optimizing the OTE procedure.

\subsection{PCG-style OT Extension} 
\label{subsec:OTE}

\begin{table}[!tb]
    \centering
    \caption{Notation used in the paper.}\label{tab:notation}
    \resizebox{1.0\linewidth}{!}{
        \begin{tabular}{cl}
        \toprule
        Notation & Explanation \\
        \midrule
        $\lambda$ & Security parameter, e.g., $\lambda=128$ \\
        $\ell$ & The length of outputs of one GGM tree\\
        $n$ & The length of outputs of $\mathrm{\Pi}_{\mathrm{SPCOT}}$\\
        $t$ & The number of generation times of GGM tree in $\mathrm{\Pi}_{\mathrm{SPCOT}}$\\
        $k$ & The length of pre-generated COT correlation in $\mathrm{\Pi}_{\mathrm{LPN}}$\\
        $s_{j}^{i}$ & The $j$-th node at the $i$-th level of the GGM tree\\
        $K_{0}^{i}$, $K_{1}^{i}$ & the XOR sum of the even or odd nodes at the $i$-th level\\
        \bottomrule
        \end{tabular}
    }
\end{table}

In hybrid HE/MPC-based PPML applications, a large number of COT correlations are required to support online non-linear function evaluation~\cite{rathee2020cryptflow2,rathee2021sirnn,huang2022cheetah,hao2022iron,pang2024bolt,lu2024bumblebee,srinivasan2019delphi}. OT Extension (OTE) is a technique that enables the generation of a large number of COT correlations. 
Up to now, there are three types of OTE: IKNP~\cite{C:IKNP03,CCS:ALSZ13,C:KelOrsSch15,C:Roy22}, PCG~\cite{CCS:BCGI18,C:BCGIKS19,CCS:BCGIKR19,CCS:YWLZW20,C:CouRinRag21,C:BCGIKR22,EC:GYWZXZ23,C:RagRinTan23} and PCF~\cite{FOCS:BCGIKS20,C:BCGIKR22,PKC:CouDuc23,CDDKS24}.
Among these, the PCG-style OTE stands out due to two key properties:  \ding{172} It reduces communication complexity from linear to sub-linear, which helps alleviate the communication bottleneck---a major challenge for both algorithmic scalability and hardware implementation. This low communication overhead has led to the widespread adoption of PCG-style OTE in state-of-the-art PPML frameworks~\cite{huang2022cheetah,pang2024bolt,lu2024bumblebee}. \ding{173} Its computation overhead scales by over $4.3\times$, creating a demand for hardware acceleration. The shift in the performance bottleneck from communication to computation in PCG-style OTE motivates our design of specialized hardware tailored to this protocol.

Except for the initialization phase that runs only once, a PCG-style OTE protocol consists of two phases:

\begin{itemize}
    \item {\bf Interactive SPCOT sub-protocol:} A sender, with the input of a random $\Delta$, and a receiver, with no input, jointly execute the SPCOT subprotocol. At the end of the sub-protocol execution, the sender obtains a random vector ${\bf w}$, the receiver outputs two vectors ${\bf u}$ and ${\bf v}$, such that $\mathbf{w} = \mathbf{v} \xor \mathbf{u} \Delta$). Here, $\Delta$ is a $\lambda$-bit block, typically with $\lambda=128$, ${\bf u}$ is a one-hot vector, where each element is a bit. Additionally, ${\bf u}$ and ${\bf v}$ refer to vectors, with each element representing a block.
    
    \item {\bf Local LPN encoding:} The sender and receiver locally expand $k$ COT correlations and $n$ SPCOT outputs to $n$ COT correlations by locally matrix-vector multiplication. 
\end{itemize}

\subsubsection{SPCOT}
In this function, the sender and receiver need to generate GGM tree~\cite{GGM86} using a double-length PRG $G$.
On CPUs, due to the availability of specialized AES-NI instruction, $G$ is usually implemented with AES operations as below:
$$\{s_{2j}^{i+1},s_{2j+1}^{i+1}\}=G(s^i_j)=\{{\mathrm{AES}}_{k_0}(s^i_j)\oplus s^i_j,{\mathrm{AES}}_{k_1}(s^i_j)\oplus s^i_j\}$$

In addition to the AES encryption, $G$ requires the output of AES to be XORed with its input. However, as  this final XOR step has a negligible computational overhead and can be integrated into the AES unit, we omit it in the subsequent discussion for simplicity.

A simplified example of $4$-node GGM tree generation is shown in Figure~\ref{fig:vole_process}(b), which consists of the following steps. 
 \textbf{\ding{172} Local Expansion}: The sender samples a random seed $s_0^0$ and then, computes the GGM tree locally from $s_{0}^{0}$,
which consumes $6$ AES calls. \ding{173} \textbf{Key Generation}: The sender computes $K^i_0$ (resp., $K^i_1$) as the XOR of the values at the even (resp., odd) nodes on the $i$-th level. The sender uses $K^i_0$ and $K^i_1$ as inputs for a 1-out-of-2 OT, allowing the receiver to make a selection based on the randomly sampled $\alpha$. 
\ding{174} \textbf{Tree Reconstruction}: The receiver uses the selected message $K_{\overline{\alpha_{i}}}^{i}$ to reconstruct all the leaf nodes, except for the $\alpha$-th leaf node, where $\overline{\alpha_{i}}$ denotes the negation of $i$-th bit of $\alpha$. 
In this example, for the first level, the receiver obtains $s^1_1$, which can then be extended to $s^2_2$ and $s^2_3$ by calling a double-length PRG locally. For the second level, the receiver consumes OT to obtain the Xor sum of $s_2^0$ and $s_2^2$. Since the receiver already possesses $s^2_2$, he can recover $s^2_0$. At this stage, only the $\alpha$-th leaf node remains unknown. \ding{175} \textbf{Node Recovery}: The sender transmits the XOR sum of all leaf nodes along with $\Delta$ to the receiver. The receiver then uses the nodes reconstructed in \ding{174} to compute the value of the $\alpha$-th leaf node, , which is given by $s^2_1 \oplus \Delta$. 

Among all the steps, \textbf{Local Expansion} incurs the highest computational overhead for the sender, while \textbf{Tree Reconstruction} imposes the greatest computational burden on the receiver. These two steps share a similar process and involve numerous AES calls to construct a binary GGM tree. Therefore, our SPCOT optimization strategy primarily focuses on optimizing these steps.



\subsubsection{LPN}
\label{subsec:lpn_background}
Before performing the LPN encoding\cite{blum2003noise}, the sender and receiver have generated $k$ COT correlations, where $\mathbf{r} = \mathbf{e}  \Delta \oplus \mathbf{s}$, and $n$ SPCOT outputs, where $\mathbf{w} = \mathbf{u}  \Delta \oplus \mathbf{v}$. Next, the sender and receiver perform a local matrix-vector multiplication: the sender computes $\mathbf{z} = \mathbf{r}  \mathbf{A} \oplus \mathbf{w}$, while the receiver computes $\mathbf{x} = \mathbf{e}  \mathbf{A} \oplus \mathbf{u}$ and $\mathbf{y} = \mathbf{s}  \mathbf{A} \oplus \mathbf{v}$. Here, $\mathbf{r}$, $\mathbf{s}$, $\mathbf{w}$, and $\mathbf{v}$ are vectors, each element of which is a 128-bit block; $\mathbf{u}$ and $\mathbf{e}$ are bit vectors; and $\mathbf{A}$ is a bit matrix. Because all elements of index matrix $\mathbf{A}$ are in the field $\{0, 1\}$, the matrix-vector multiplication with $\mathbf{A}$ can be formulated as a random-memory-access problem. In the baseline parameter set, each row contains 10 non-zero elements, indicating we need to compute the XOR sum of 10 randomly accessed values to obtain one result.

The LPN protocol has the following two main properties that can benefit the hardware design. Firstly, the matrix $\mathbf{A}$ only needs to be generated once and remain fixed for different rounds of LPN computation. This not only indicates the generation of $\mathbf{A}$ can be ignored but also provides the opportunity to sort $\mathbf{A}$ to transform the random memory access into a more regular access pattern to reduce the DRAM access latency. Secondly, in the LPN protocol, the generation of $\mathbf{e}$ and $\mathbf{A}$ is independent of the SPCOT output $\mathbf{u}$, providing the opportunity to overlap LPN with SPCOT for further efficiency improvement.

%% file: docs/3-motivation_v2.tex
\begin{figure*}[!tb]
    \centering
    \includegraphics[width=0.9\linewidth]{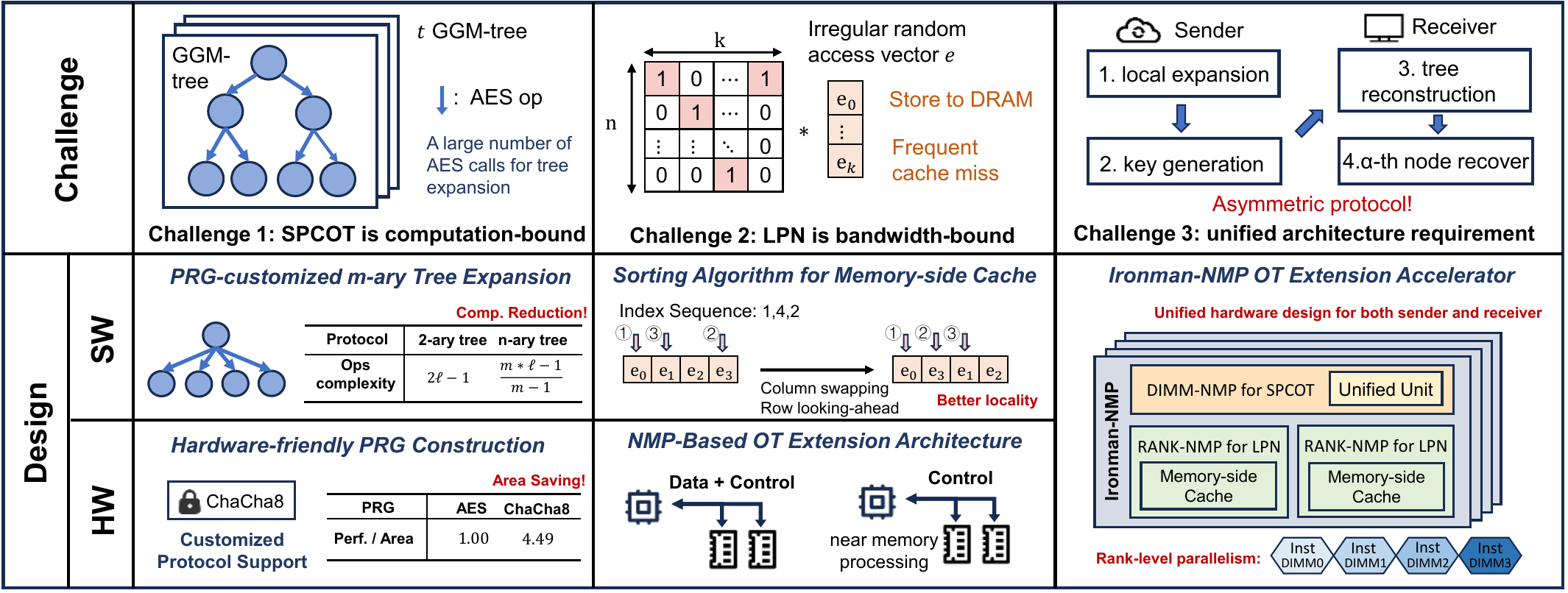}
    \caption{ The challenges of accelerating PCG-style OT extension and the corresponding HW/SW design overview.
    }
    \label{fig:method_overview}
    \Description{Methodology Overview}
\end{figure*}

\section{Motivation: Identifying Bottleneck in PCG-style OT Extension}

To accelerate the PCG-style OTE, we must address three key challenges: (1) the computation bottleneck of SPCOT, (2) the memory bandwidth bottleneck of LPN, and (3) the requirement of a unified architecture for the sender and receiver. We have the following observations concerning these challenges.


\subsection{SPCOT Operation Analysis}
\label{subsec:MPCOT_analysis}



\begin{figure}[!tb]
    \centering
    \includegraphics[width=0.95\linewidth]{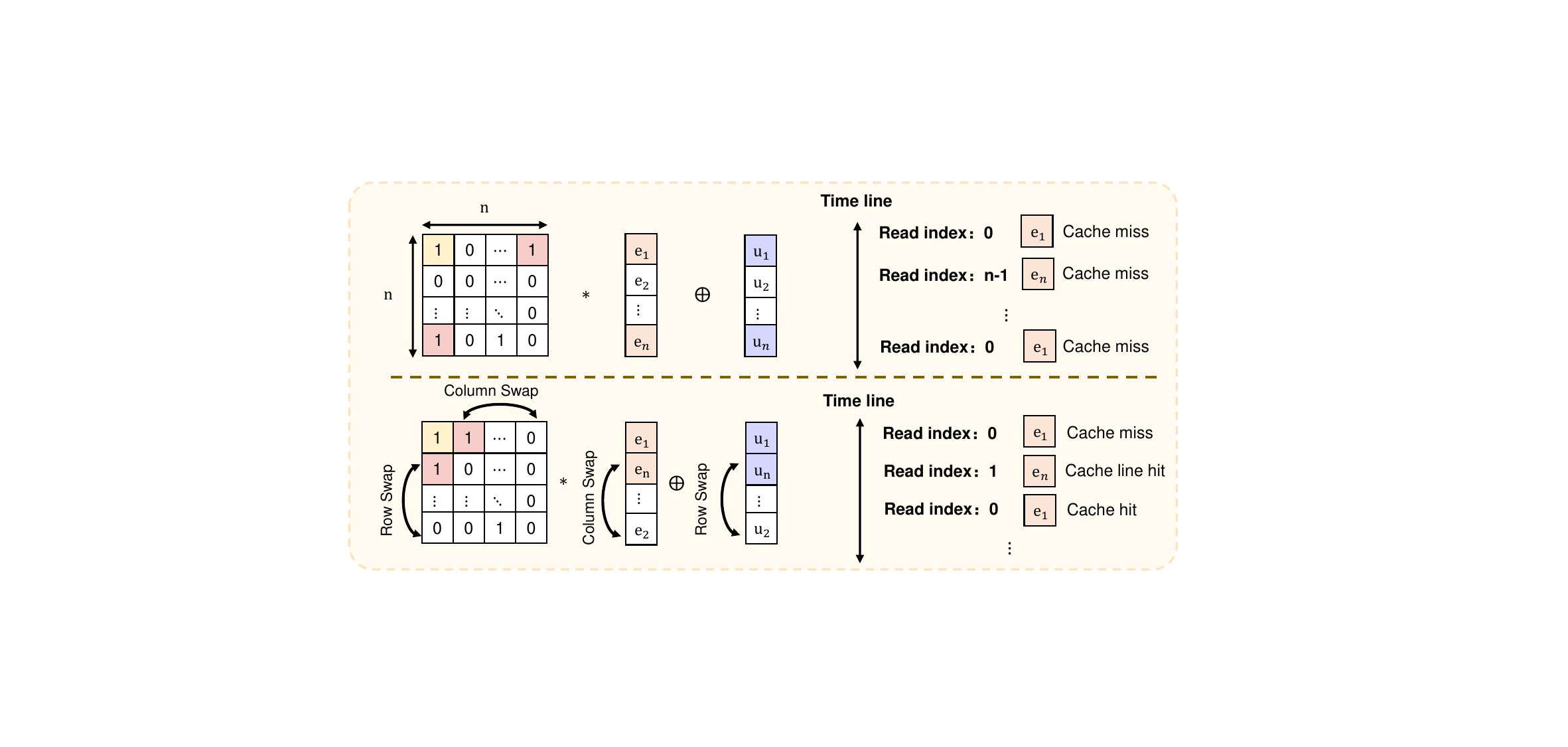}
    \caption{A simple example demonstrates how column and row swaps can enhance temporal and spatial data locality.}
    \label{fig:sort_example}
    \Description{An example of sorting}
    \vspace{-5pt}
\end{figure}

\textbf{Observation 1: SPCOT operations exhibit high computational demands and reducing operation calls can be a better solution.} To accelerate SPCOT, we can either increase the number of PRG (e.g., AES) cores or reduce the total number of PRG calls. Increasing the number of PRG cores incurs more area overhead. To reduce the number of PRG calls, we have the following opportunities: 

\textbf{(1) 2-ary tree expansion is not an optimal algorithm choice.} Besides using 2-ary tree for GGM tree construction, we can also leverage m-ary tree expansion \cite{CCS:SGRR19}. To generate a tree with $\ell$ leaf nodes, m-ary tree expansion can reduce the number of AES calls from $2\ell - 1$ for 2-ary tree to $\frac{m \cdot \ell - 1}{m - 1}$ AES calls, thereby improving the generation efficiency.

\begin{table}[!tb]
    \centering
    \caption{PRGs Comparison. Perf./Area Ratio and Power/Block Ratio are normalized to AES.}
    \label{tab:PRG_comparison}
    \scalebox{0.85}{
    \vspace{3pt}
    \begin{tabular}{c|c|c|c|c|c}
    \toprule
    \multirow{2}{*}{PRG} & \multirow{2}{*}{\makecell[c]{Output size \\(bit)}} & \multirow{2}{*}{\makecell[c]{Area\\ (${mm}^2$) }}  & \multirow{2}{*}{\makecell[c]{Perf./Area \\Ratios}} & \multirow{2}{*}{\makecell[c]{Power\\ ($mW$) }} & \multirow{2}{*}{\makecell[c]{Power/Block \\Ratios}}   \\ 
    & & & & & \\
    \midrule
    AES-128 & 128 & 0.233  & 1 & 35.05 & 1 \\
    \midrule
    ChaCha8 & 512 & 0.215 & 4.491 & 45.34 & 3.092 \\
    \bottomrule
    \end{tabular}}
\end{table}



\textbf{(2) AES-based PRG is not optimal for customized protocol.} 
In customized hardware scenarios, ChaCha-based PRGs exhibit better properties.  As shown in Table~\ref{tab:PRG_comparison}, we evaluate different fully-pipelined PRGs from three perspectives: \textbf{\ding{172} Area}: ChaCha8 is a suitable choice for area efficiency, occupying $0.215{mm}^2$, compared to $0.233{mm}^2$ for AES. \textbf{\ding{173} Power}: ChaCha8 consumes 45mW, compared to AES's 32mW. However, ChaCha8 produces 4 blocks per cycle (512 bits) while AES produces only one block per cycle (128 bits). Therefore, ChaCha8 is more power-efficient per block. \textbf{\ding{174} Security}: The 128-bit security level is sufficient for most of real-word applications. According to~\cite{aumasson2019too}, 7-round ChaCha provides around 248-bit security. These advantages have been well recognized in the literature~\cite{lam2023gpu,xiaolin2025pota}. However, many of these works overlook a key property of ChaCha-based PRGs: their ability to generate longer output lengths. This feature is particularly suited for m-ary tree expansion and can be leveraged to further reduce the number of PRG calls.

\subsection{LPN operation Analysis}
\label{subsec:lpn_analysis}

\textbf{Observation 2: LPN operation requires high memory bandwidth due to random memory access and simple computation.}
As LPN is memory bandwidth-bound, we highlight two factors to overcome the bandwidth limitation of LPN operations as follows.

    

\textbf{(1) The compute-to-memory access ratio of LPN is relatively low, which intensifies the problem of bandwidth bound.} 
We use the LPN computation on the sender's side as an example. Following Section~\ref{subsec:OTE}, the sender needs to locally compute $\mathbf{z} = \mathbf{r} \cdot \mathbf{A} \oplus \mathbf{w}$. All elements of $\mathbf{A}$ are binary, and each row of $\mathbf{A}$ contains exactly 10 non-zero elements. This implies that reading a 128-bit element requires only a single XOR operation, resulting in a very low computational memory access ratio. 
Furthermore, when the OT output size exceeds $2^{24}$, storing $\mathbf{r}$ and $\mathbf{A}$ consumes more than 900MB, making it challenging to leverage the cache for improved bandwidth.
Given the low \textbf{compute-to-memory acces ratio}, NMP provides an opportunity to exploit the high internal bandwidth of DRAM while supporting simple operations, making it particularly well-suited for LPN due to its minimal computational requirements.

\textbf{(2) Sorting the index matrix can further improve data locality, effectively increasing memory bandwidth.} LPN suffers from low memory bandwidth due to irregular memory access patterns. Given that LPN can be formulated as a matrix-vector multiplication, row and column swaps can be applied to cluster non-zero elements into regular structures without affecting the output. Figure~\ref{fig:sort_example} illustrates a simple example showing how row and column swaps can enhance data locality. Furthermore, LPN has the advantage that the index matrix can be fixed, and hence, the index sorting can be conducted once during compile time without causing online latency overhead.


\subsection{Unified Architecture}

\textbf{Observation 3: the sender and receiver share similar computation except for the XOR sum computation in SPCOT.} 
During Key Generation, the sender needs to compute the XOR sums of values at both the even and odd nodes. In contrast, during Tree Reconstruction, the receiver selects from the even and odd XOR sums from the sender obliviously and uses the received XOR sum to assist the downstream GGM tree construction. Hence, the sender and receiver both require XOR trees for GGM tree construction and the main difference comes from the dataflow, which alleviates the hardware cost for unified support.


\subsection{Proposal optimization Overview}
Figure~\ref{fig:method_overview} illustrates the challenges of accelerating OT extension and our proposed OTE accelerator \sysname.
\sysname~features an NMP-based architecture and co-designs the algorithm and hardware to boost the efficiency of PCG-style OTE.
For SPCOT, we propose a hardware-awared m-ary tree expansion optimization. For LPN, we design an NMP architecture equipped with a memory-side cache to leverage the sorted index matrix. We also introduce a unified architecture to support both the sender's and receiver's protocols.

%% file: new_method_2/spcot.tex



\begin{figure}[!tb]
    \centering
    \includegraphics[width=0.85\linewidth]{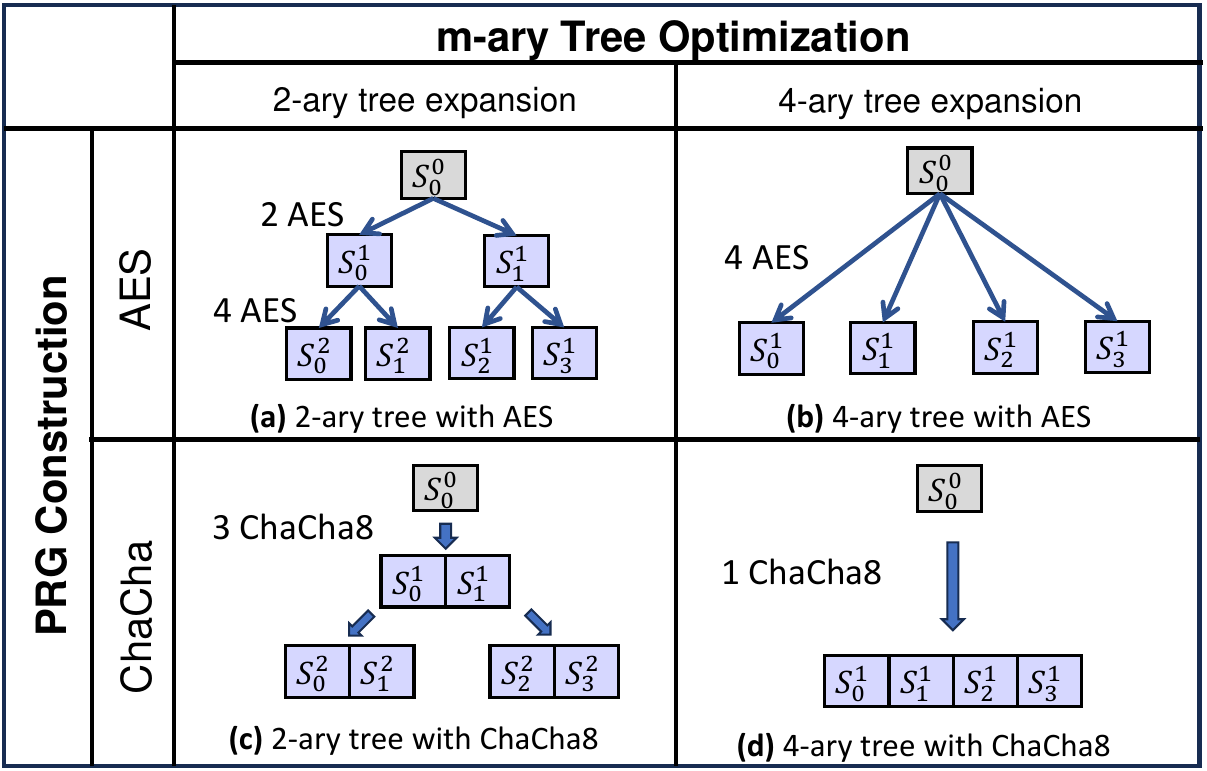}
    \caption{A four-node example of the SPCOT optimization framework.
    }
    \label{fig:m-ary_tree}
    \Description{A four-node example of the SPCOT optimization}
\end{figure}

\section{Hardware-Awared M-ary Tree Expansion}
\label{subsec:hf_prg}


To accelerate SPCOT operation, we introduce Hardware-Awared M-ary Tree Expansion that combines m-ary tree expansion with ChaCha-based PRGs can yield cooperative benefits, resulting in a performance improvement that exceeds the sum of its parts. Additionally, we utilize a GGM-tree to generate $(m-1)$-out-of-$m$ OT, supporting m-ary tree expansion with sublinear complexity in OT correlation consumption. This framework reduces operation calls $6\times$ while maintaining the same area consumption on-chip. Furthermore, it can be implemented in any hardware environment without AES-NI to achieve improved OTE performance.

\subsection{PRG-customized m-ary Tree Expansion}
\label{subsec:m-ary_tree}



Given that ChaCha-based PRGs generates longer output lengths than other constructions and the maximum operation reduction rate of solely m-ary tree expansion is inherently limited, we integrate ChaCha-based PRG with m-ary tree expansion to achieve greater reduction improvements. 

As shown in Figure~\ref{fig:m-ary_tree}(a) and (b), in the baseline protocol, two AES operations with different keys are used as a double-length PRG, requiring a total of six AES operations to generate four leaf nodes. In contrast, employing AES with four unique keys as a quadruple-length PRG requires only four AES operations.
More generally, to obtain $\ell$ leaf nodes, a m-ary tree expansion requires only $\frac{m \cdot \ell - 1}{m - 1}$ AES operations, compared to $2\ell-1$ operations required by a 2-ary tree expansion. However, as $m$ continues to increase, the operation reduction rate asymptotically approaches $2$, highlighting the limited potential for further improvement. 

To overcome the reduction limitation, we replace the AES-based design with a ChaCha-based PRG. The ChaCha-based PRG has a longer output, allowing it to implement double-length and quadruple-length PRGs with a single operation call, as shown in Figure~\ref{fig:m-ary_tree}(c) and (d). 
By employing a ChaCha-based PRG, the number of operation calls is reduced by 50\% and 75\% for a 2-ary and 4-ary tree expansion, respectively.

\begin{figure}[!tb]
    \centering
    \includegraphics[width=1.0\linewidth]{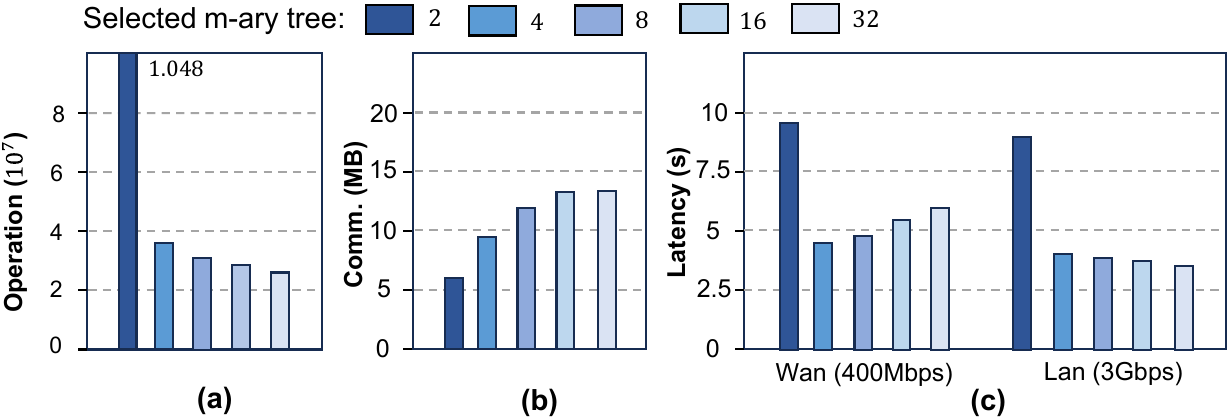}
    \caption{ m-ary Tree Ops, Communication cost, and latency Comparison under Wide Area Network (WAN) and Local Area Network (LAN) settings.}
    \label{fig:m-ary_choice}
    \Description{m-ary Tree Expansion ops and Comm.}
\end{figure}

An important problem that remains to be solved is determining the optimal choice of m-ary tree. Because ChaCha-based PRGs can generate up to four blocks at most by one operation, meaning that as $m$ increases, the benefits of m-ary tree expansion in reducing operations become limited. This trend is illustrated in Figure~\ref{fig:m-ary_choice}(a): while a 4-ary tree expansion achieves a $2.99\times$ reduction in operations compared to 2-ary expansion with ChaCha, a 32-ary tree expansion achieves only a $3.86\times$ reduction compared to 2-ary. 
Additionally, as demonstrated in Figure~\ref{fig:m-ary_choice}(b) and (c), we observe that as $m$ increases, the online communication demands and latency also increase. This indicates that selecting a larger $m$-ary tree negatively impacts latency in bandwidth-limited scenarios. Consequently, we select 4-ary expansion as our optimized algorithm, as it achieves significant operation reduction with minimal communication overhead, making it well-suited for bandwidth-limited scenarios.

\begin{figure}[!tb]
    \centering
    \includegraphics[width=0.9\linewidth]{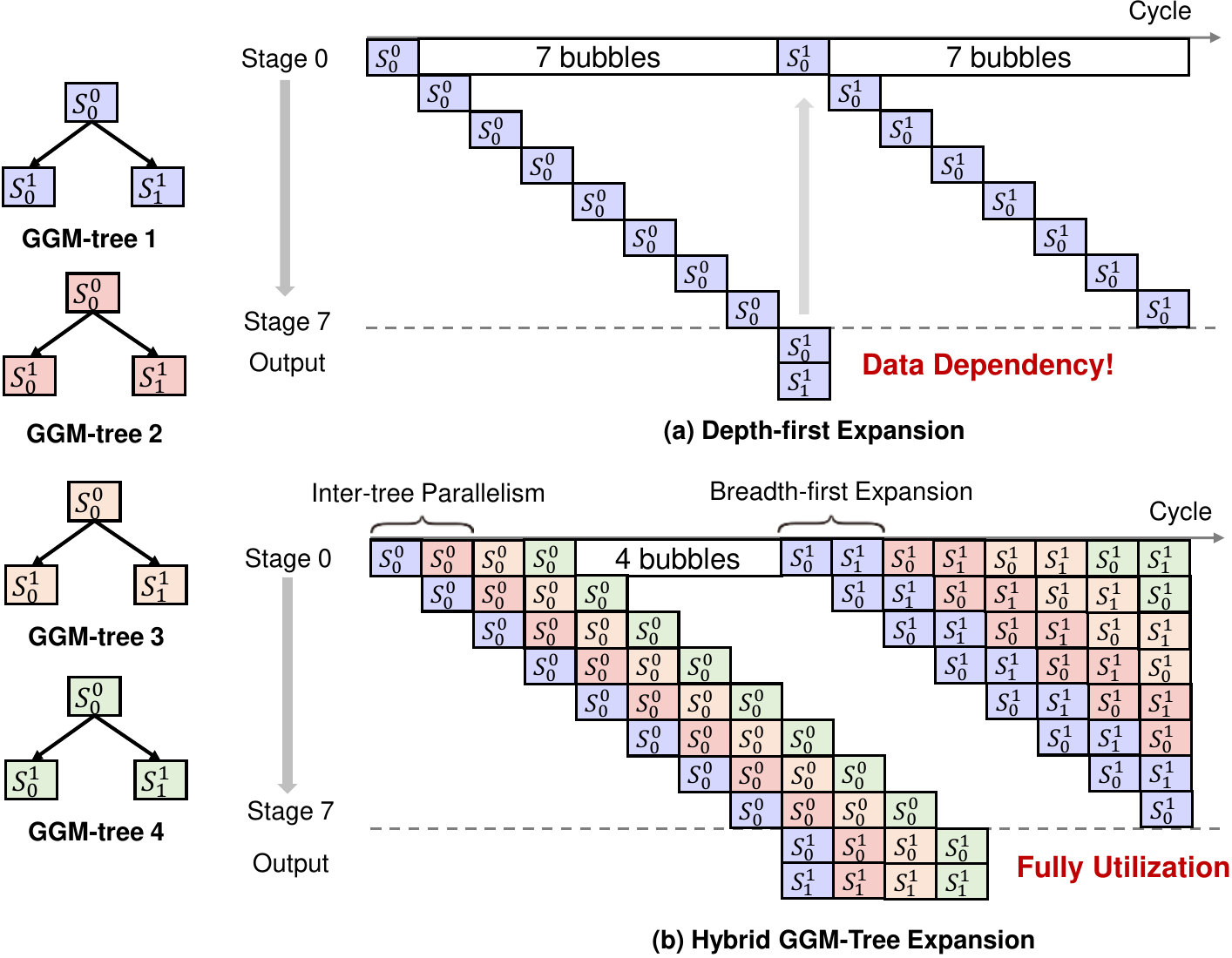}
    \caption{(a) Depth-first Expansion; (b) Pipeline-based Tree Expansion Strategy.}
    \label{fig:pipeline}
    \Description{Depth-first Expansion and Pipeline-based Tree Expansion Strategy}
\end{figure}

\subsection{(m-1)-out-of-m OT Generation} 
However, m-ary tree expansion has a side effect: it requires $(m-1)$-out-of-$m$ OT instead of the standard 1-out-of-2 OT,  resulting in additional computational overhead due to the more complex generation protocol. 
The definition of $(m-1)$-out-of-$m$ OT is similar to 1-out-of-2 OT described in Section~\ref{subsec:ot}. In this case, the sender inputs $m$ messages $m_0,m_1,\dots,m_m$  while the receiver inputs an index $\alpha \in [m]$. After the protocol,  the receiver obtains all messages except $m_{\alpha}$. 

The naive method directly that using $(m-1)*log_2^m$ 1-out-of-2 OT to implement a $(m-1)$-out-of-$m$ OT has significant drawbacks. It results in a waste of OT correlations, leading to increased executions of OTE and requires customized hardware to support the protocol. We observe that a $(m-1)$-out-of-$m$ OT can be efficiently generated using an $m$-node GGM tree. As an example, consider $m=4$. The sender and receiver generate a 4-node GGM-tree. According to the definition of the GGM tree, the sender obtains four random strings $r_0$, $r_1$, $r_2$ and $r_3$ , while the receiver obtains three of these four strings, excluding $r_{\alpha}$, where $\alpha \in \{0,1,2,3\}$. Using the same OT correlation method, the sender XORs the input messages with the outputs of the GGM tree and sends them to the receiver. The receiver then decrypts $3$ messages, excluding $m_{\alpha}$. This approach only consumes $log_2^m$ OT correlations to generate a $(m-1)$-out-of-$m$ OT and follows the same procedure as SPCOT, meaning there is no need for additional hardware units.

\subsection{GGM-tree Expansion Schedule}
\label{subsec:ggm-tree_traversal}

From the hardware implementation perspective, there are two expansion strategies we can adopt: depth-first and breadth-first expansion. In depth-first expansion, the ChaCha8 subsequent executions have to wait for 8 cycles to commence because of data dependency, as we present in Figure~\ref{fig:pipeline}(a). Conversely, breadth-first expansion allows for full utilization of the 8-stage pipeline. However, this approach suffers from memory overhead, requiring a buffer size of $O(\ell)$, compared to the $O(\log{\ell})$ buffer size required by depth-first expansion, as analyzed in \cite{lam2023gpu}. Additionally, in the PCG-style OTE, each leaf node can be combined with an LPN output to achieve the desired OT. The breadth-first expansion delays the readiness of the leaf nodes, further increasing the memory overhead associated with the LPN operations.

Based on the analysis above, we propose the Hybrid GGM-tree Expansion Strategy, which builds on a depth-first approach to reduce memory consumption. Figure~\ref{fig:pipeline}(b) presents an example for generating four two-level GGM trees. The key idea is to mitigate pipeline idling in the ChaCha core during child node computation by employing breadth-first expansion and inter-tree parallelism. When the current level contains sufficient nodes, breadth-first expansion is used to compute other nodes at the same level concurrently. Conversely, when there are insufficient nodes to fully utilize the pipeline, inter-tree parallelism is applied to eliminate pipeline bubbles. By implementing the Hybrid GGM-tree Expansion Strategy, $100\%$ utilization of the ChaCha cores can be achieved.

%% file: new_method_2/lpn.tex
\begin{figure*}[!tb]
    \centering
    \includegraphics[width=0.88\linewidth]{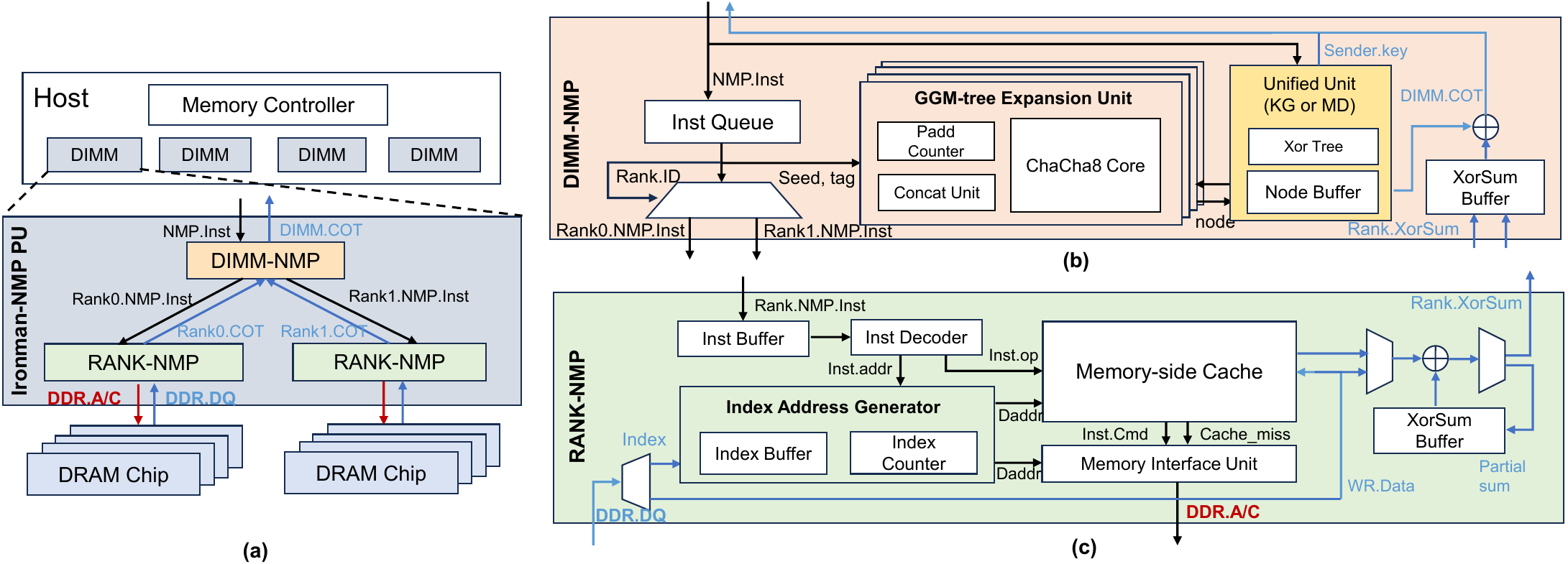}
    \caption{(a) Ironman-NMP architecture overview. (b) DIMM module. (c) Rank module.
    }
    \label{fig:NMP}
    \Description{NMP Architecture Overview}
\end{figure*}

\section{Ironman-NMP Architecture}

\subsection{Accelerator Design}
\label{subsec:arch_design}
Many works leverage near-memory processing to provide higher bandwidth for their target applications, such as RecNMP~\cite{ke2020recnmp}, DIMMining~\cite{dai2022dimmining}, and MetaNMP~\cite{chen2023metanmp}. Considering the memory-bound nature and the sparse, irregular access patterns of LPN operations, we propose Ironman, a near-memory processing solution specifically designed to accelerate LPN operations. Ironman exploits DRAM rank-level parallelism by performing computations directly and locally on data fetched from activated ranks, thereby achieving greater bandwidth. Additionally, In order to achieving end-to-end COT generation, Ironman efficiently implements the SPCOT operation using minimal hardware resources, 
Ironman is located on the buffer chip of the DIMM. Each buffer chip contains an Ironman-NMP processing unit (PU) made up of a DIMM-NMP module and two Rank-NMP modules, as illustrated in Figure~\ref{fig:NMP}(a). 

At the beginning of PCG-style OTE, an initialization is required for both sender and receiver. The host evenly partitions the index matrix $\mathbf{A}$ into multiple segments and distributes them across the ranks. We then perform row-wise partitioning of matrix $\mathbf{A}$, to enable parallelism across ranks. In Ferret~\cite{CCS:YWLZW20}, each row of the matrix contains exactly 10 non-zero elements, ensuring that all ranks have an equal workload. Additionally, it broadcasts the pre-generated vectors ($\mathbf{r}$ for the sender and $\mathbf{s}$ and $\mathbf{e}$ for the receiver) to all ranks. This step decouples the operation of each rank.

In the next step, the DIMM module receives an NMP instruction (NMP-Inst) from the memory controller and forwards it to the appropriate Rank module based on the rank address. The Rank modules execute the NMP-Inst to compute the XOR sum of randomly accessed numbers concurrently. The DIMM then performs the SPCOT operation and computes the XOR of the SPCOT outputs with the XOR sum from parallel ranks to produce the final result (DIMM.COT).
Moreover, in our design, the SPCOT and LPN operations are decoupled, allowing us to overlap these two operations to enhance performance. In the following parts, we will describe the design details of the DIMM-NMP and Rank-NMP modules.

\subsubsection{DIMM-Module}
Figure~\ref{fig:NMP}(b) shows the architecture of the DIMM module, which performs three key functions: dispatching NMP instructions from the memory controller, executing the SPCOT operation, and computing COT correlations. The DIMM-module consists of two key units: the ChaCha8 Core and the Unified Unit. The ChaCha8 Core is responsible for generating the GGM tree using the Hardware-Awared m-ary Tree Expansion algorithm. The Unified Unit, implemented as an XOR tree, supports both sender and receiver protocols to meet the requirements of MPC scenarios. The details of Unified Unit is described in Section\ref{subsec:unified}.

\subsubsection{Rank-Module}
Figure~\ref{fig:NMP}(c) shows the architecture of rank module, which also performs three key functions: translating the NMP-Inst into DDR command/address (C/A), managing memory-side caching and computing the XOR sum of randomly accessed numbers for LPN operation. The Rank-module also consists of two key units: the Index Address Generator and the Memory-side Cache. The Index Address Generator is responsible for computing the required addresses, thereby eliminating the need to frequently return indices to the memory controller.
Additionally, we equip the Rank-module with a memory-side cache and implement a sorting algorithm to further improve data locality, addressing the sparse and irregular memory access pattern in LPN operations. The concrete method is presented in Section\ref{subsec:cache}.


\subsubsection{Design Details}
\label{subsubsec:nmp_details}
Regarding the offloading cost, the main overhead comes from sending a large number of COT correlations back to the CPU. For example, the first layer of ResNet-50 requires over $4 \times 10^7$ COT correlations, totaling over 500 MB, which takes 8.1 ms to transmit (assuming 76.8 GB/s DDR4 bandwidth). However, in \sysname, COT correlations are generated sequentially, meaning there is no need to wait for all COT correlations to be generated before starting to send them back to the CPU. Instead, we can overlap their generation and transmission. As a result, the offloading cost becomes negligible.

Regarding compatibility with existing commercial NMP-compatible hardware, systems such as UPMEM and Samsung HBM-PIM can be deployed for \sysname, but they present certain challenges. UPMEM uses general-purpose cores, which may not provide sufficient performance for ChaCha. In contrast, Samsung HBM-PIM integrates specialized AI computation logic, but it does not fully meet the specific requirements of \sysname. This would require replacing the existing hardware with our custom ASIC, although the overall architecture remains similar.

\subsection{Unified Architecture Design}
\label{subsec:unified}
\begin{figure}[!tb]
    \centering
    \includegraphics[width=1\linewidth]{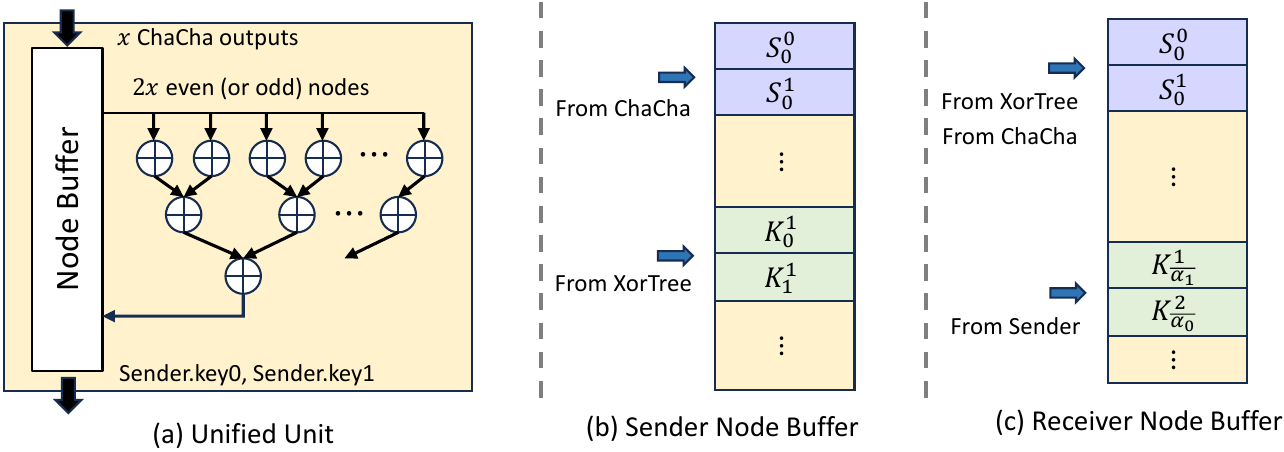}
    \caption{(a) The Unified Unit operates either as a Key Generator for the sender or as a Message Decoder for the receiver. (b) Node Buffer under the sender protocol. (c) Node Buffer under the receiver protocol.
    }
    \label{fig:unified}
    \Description{Unified Unit}
\end{figure}

In MPC scenarios, both linear and non-linear functions can benefit from exchanging the roles of sender and receiver. For example, as demonstrated by PrivQuant \cite{xu2024privquant}, an OT-based matrix multiplication protocol can achieve better communication overhead reduction by swapping the roles of sender and receiver between the server and the client. Additionally, in CrypTFlow2 \cite{rathee2020cryptflow2}, the Multiplexer and Beaver-Triple Generation protocols require both parties to act as sender and receiver, necessitating the ability to execute the OTE protocol from either perspective. Finally, role switching also enhances multithreaded parallelism, allowing the server or receiver to prepare messages while waiting for incoming ones. 
In conclusion, supporting both sender and receiver protocols can benefit protocol design, motivating us to create a unified architecture.
Additionally, While the CPU baseline is unified, it has limited performance. Naive ASIC implementations for sender and receiver protocols can achieve high performance but result in a non-unified architecture due to differing operations on each side of the protocol. Our design combines the best of both approaches, offering both unification and high performance.


Therefore, to support both sender and receiver protocols, we implement a Unified Unit that functions as either a Key Generator or a Message Decoder, depending on the host's role (Figure~\ref{fig:unified} (a)). The size of the XOR tree is determined by the number of ChaCha cores per DIMM module; with $x$ ChaCha cores producing 512-bit outputs, a $2x$-node XOR tree computes partial sums of even or odd nodes. In the sender protocol, the unit computes sums of both even and odd nodes and sends the results, requiring two executions of the XOR tree. In the receiver protocol, it computes only one sum (either even or odd nodes) and stores the recovered node in the Node Buffer. This design allows seamless role switching using the same hardware. Figure~\ref{fig:unified}(b) and (c) show the differences in the Node Buffer between the sender and receiver. The Node Buffer stores both nodes and keys. For the sender, nodes are obtained from the ChaCha Core output, while for the receiver, nodes are either generated by the ChaCha Core or recovered using the XOR Tree. The node size is determined by the GGM tree height, calculated as $\log_4\ell$. For the key portion, the sender generates both even and odd XOR sums, while the receiver only needs one.

\begin{figure}[!tb]
    \centering
    \includegraphics[width=0.90\linewidth]{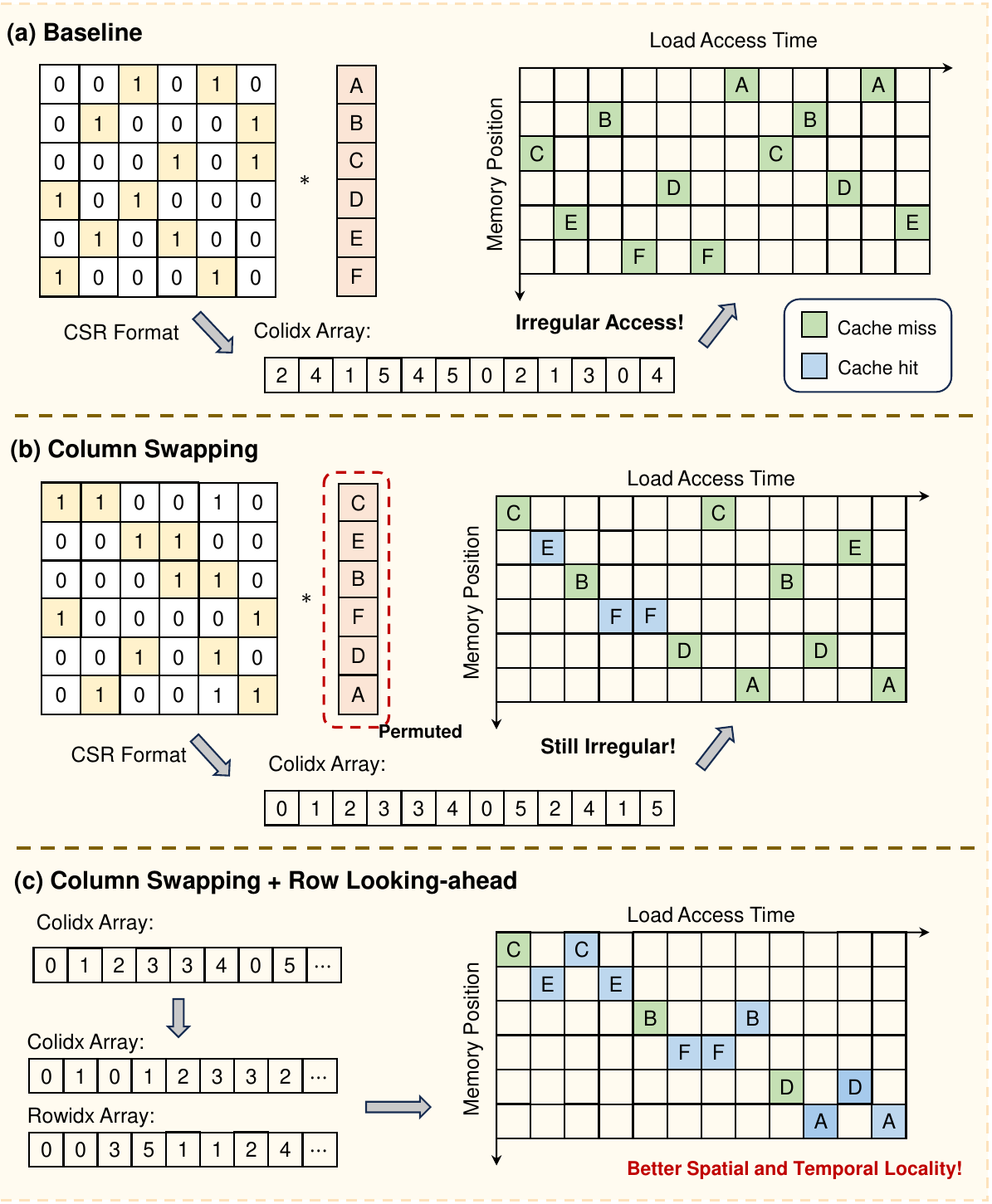}
    \caption{A toy example demonstrating how Column Swapping and Row Look-ahead work is presented. In this example, both the cache size and cache line are set to 2.
    }
    \label{fig:sorting_algorithm}
    \Description{Sorting Algorithm}
\end{figure}


\subsection{Index Sorting Algorithm for Memory-side Cache}
\label{subsec:cache}
LPN can be formulated as a Sparse Matrix-Vector Multiplication (SpMV), typically executed using the Compressed Sparse Row (CSR) format. The CSR format encodes a sparse matrix with three arrays:  $\mathit{Value}$ as the nonzero values, $\mathit{Colidx}$ as the column indices of nonzero values, and $\mathit{Rowptr}$ to store the pointers of each row’s first element in the value and $\mathit{Colidx}$ array. In this case because the property of LPN that each row only has 10 non-zero elements and element are in the field $\{0,1\}$, Therefore, we only need a $\mathit{Colidx}$ to store the sparse matrix. Due to the irregular memory access pattern and poor data locality of LPN, this operation results in significant memory access latency, which degrades overall performance. 

A naive approach to address this problem is to utilize a cache to reduce memory access latency. However, as Figure~\ref{fig:sorting_algorithm}(a) illustrates, poor data locality leads to numerous cache misses, preventing the full utilization of the cache bandwidth. To mitigate this issue, we propose an index sorting algorithm combined with a memory-side cache to improve performance. The algorithm consists of two steps: Column Swapping and Row Look-ahead.
Figure~\ref{fig:sorting_algorithm}(b) illustrates how Column Swapping improves spatial data locality. In this example, we change the column sequence from $\mathbf{A,B,C,D,E,F}$ into $\mathbf{C,E,B,F,D,A}$. This transformation helps convert irregular access patterns into sequential access, which can take advantage of cache lines to reduce cache miss rates. However, our experiments show that using Column Swapping alone achieves a maximum cache hit rate of only 20\% with a 1MB cache, which still results in significant latency for LPN operations.

To achieve further latency reduction, we introduce Row Look-ahead cooperated with Column Swapping. Figure~\ref{fig:sorting_algorithm}(c) give a simple example of how Row Look-ahead works. In previous methods, elements were read from the sparse matrix row by row. Here, we introduce $\mathit{Rowidx}$ to overcome this limitation. This allows us to pre-fetch the indices of subsequent rows and access the corresponding elements. If the elements indicated by these indices are already present in the current cache, we can access them in advance, thereby reducing the likelihood of a cache miss. 
Because the sparse matrix can remain fixed, we simulate memory-side cache behavior on the offline phase and store the sorted CSR format which have a better temporal and spatial locality.

In our NMP module, two arrays $\mathit{Rowidx}$ and $\mathit{Colidx}$ are sequentially stored in DRAM. These arrays are accessed in a streaming manner, leveraging DRAM’s high bandwidth without relying on memory-side cache for index retrieval. For each LPN operation, the NMP module reads the next index from the $\mathit{Colidx}$ array and uses it to access the corresponding entry in the error vector. Accesses to the error vector are first checked against the memory-side cache to determine whether a cache hit occurs before falling back to DRAM.
Importantly, since the $\mathit{Colidx}$ array is sorted as part of the "column swap + row forward-looking" strategy, the accessed indices to the error vector exhibit improved spatial locality. Finally, we use the $\mathit{Rowidx}$ array to indicate which read data need to be XORed together.

Finally, there are some details that need to be further explained:
\begin{itemize}
    \item  \textbf{Sorting overhead}: Due to the property of LPN, the sorted CSR format can be reused in all subsequent PCG-style OTE operations, which amortizes the sorting overhead. Additionally,  we divide the matrix into smaller blocks and perform sorting on them separately to reduce the overhead.
    \item \textbf{Vector permutation}: Column swapping does indeed require permuting the input vector. However, in this application, the input vector is indistinguishable from a random vector based on the LPN assumption\cite{blum2003noise}. Whether the input vector is permuted or not does not affect the correctness of the protocol, as the LPN assumption only requires that the sender and receiver use the same input vector
\end{itemize}



%% file: docs/6-setup.tex
\section{Experiment}
\label{sec:setup}

\begin{table}[!tb]
\renewcommand\arraystretch{1.3}
\centering
\caption{System configuration parameters.}
\label{tab:configuration}
\scalebox{0.9}{
\begin{tabular}{c|c}
\toprule
\textbf{Category}            & \textbf{Parameters} \\
\midrule
\multirow{2}{*}{Host Processor} 
                             & 24 cores, 2.2GHz, 1.5MB L1 cache, \\ 
                             & 48MB L2 cache, 71.5MB LLC cache \\ \hline

\multirow{2}{*}{DRAM Module} 
                             & DDR4-2400MHz 8GB, 64GB total size, \\ 
                             & 4 Channels × 2 DIMMs × 2 Ranks, FR-FCFS \\ \hline

\multirow{3}{*}{\makecell[c]{DRAM Timing \\ Parameters}} 
                               & tRCD=16, tCL=16, tRP=16, tRC=55, \\
                               & tRRD\_S=4, tRRD\_L=6, tFAW=26, \\
                               & tCCD\_s=4, tCCD\_L=6, tBL=4 \\
\bottomrule

\end{tabular}}
\end{table}

\begin{table}[!tb]
    \centering
    \caption{PCG-style OT-extension Parameter Sets}
    \label{tab:parameter_sets}
    \scalebox{0.9}{
    \vspace{3pt}
    \begin{tabular}{c|c|c|c|c|c}
    \toprule
     \#OTs for  & \multirow{2}{*}{$n$} & \multirow{2}{*}{$\ell$} & \multirow{2}{*}{$k$} & \multirow{2}{*}{$t$} & \multirow{2}{*}{Bit-security} \\
     output & & & & & \\
    \midrule
    $2^{20}$ & 1221516 & 4096 & 168000 & 480 & 139.8 \\ \hline
    $2^{21}$ & 2365652 & 4096 & 262000 & 600 & 141.8 \\ \hline
    $2^{22}$ & 4531924 & 8192 & 328000 & 740 & 132.3 \\ \hline
    $2^{23}$ & 8866608 & 8192 & 452000 & 1024 & 130.2 \\ \hline
    $2^{24}$ & 17262496 & 8192 & 480000 & 2100 & 135.4 \\
    \bottomrule
    \end{tabular}}
\end{table}


\textbf{Hardware Simulation.} 
Our experiments was evaluated on a 24-core Intel Xeon Gold 5220R CPU with DDR4 memory. We also implement the OTE protocol on an NVIDIA A6000 GPU.
To evaluate the performance of \sysname, we integrate Ramulator~\cite{kim2015ramulator} and ZSim~\cite{sanchez2013zsim} to build a cycle-level accurate simulator. We use a Ramulator to simulate the latency of SPCOT and LNP. The data offloading latency of NMP is simulated through ZSim. 

For the hardware area and energy consumption, we implement the Verilog and use Synopsys Design Compiler to evaluate the ChaCha8 Core under 45nm technology, and we use Cacti\cite{muralimanohar2009cacti} to evaluate the on-chip buffer.
Additionally, we use Cacti-3DD\cite{chen2012cacti} to estimate the energy consumption of DRAM devices and Cacti-IO\cite{jouppi2012cacti} for off-chip I/O energy consumption at the DIMM level.

\textbf{Parameter.} 
To demonstrate that our hardware design supports SPCOT and LPN operations across various parameter configurations, Table~\ref{tab:parameter_sets} presents the specific parameters used in our experiments. We have verified that these parameter configurations provide sufficient 128-bit security for cryptographic algorithms based on \cite{EC:LWYY24}. Conducting experiments across different parameter configurations highlights the adaptability of our algorithmic optimizations and hardware design, showcasing their suitability to various scenarios with unique parameter requirements.



%% file: docs/7-experiment.tex
\subsection{Overall PCG-style OTE Generation Speedup}\label{subsec:overall}


\begin{figure}[!tb]
    \centering
    \includegraphics[width=0.9\linewidth]{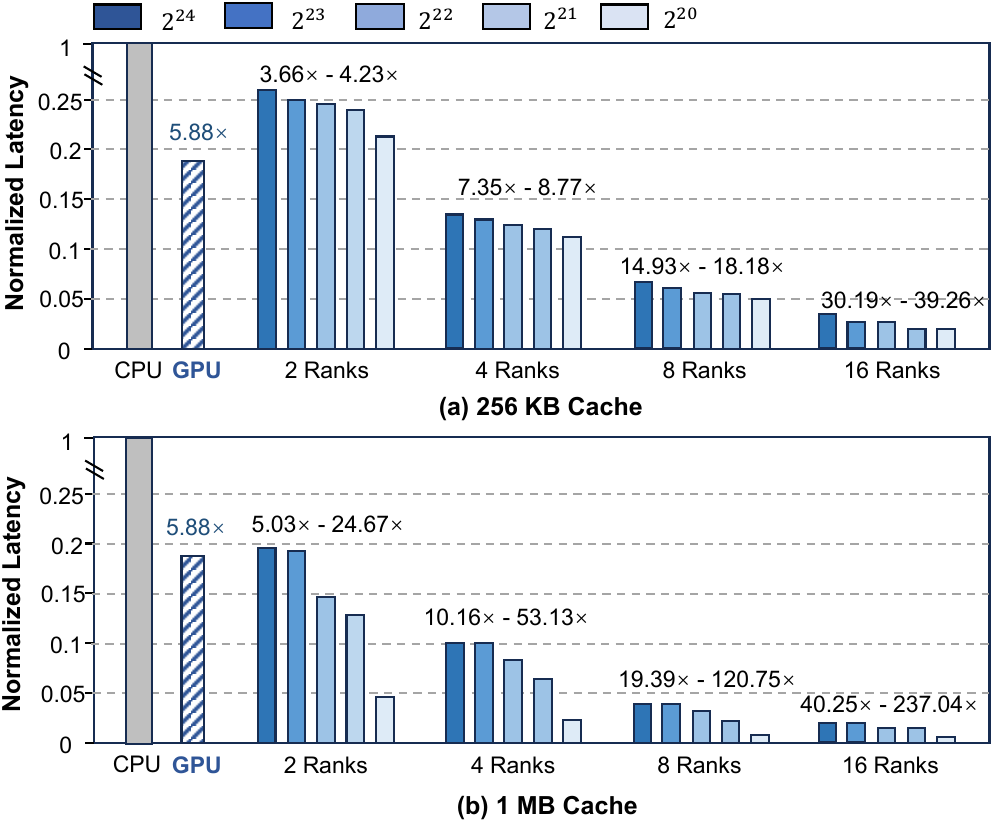}
    \caption{The Latency of OTE on CPU, GPU and Ironman with (a) 256KB cache and (b) 1MB cache, under various memory configurations and parameter sets, when $2^{25}$ OTs (latency is normalized to CPU baseline).
    }
    \label{fig:speedup}
    \Description{Speedup Result}
\end{figure}

We use Ferret~\cite{CCS:YWLZW20} as our baseline, representing the SOTA performance for PCG-style OT extension on CPU. We also implement the OTE protocol on an NVIDIA A6000 GPU for a comprehensive hardware comparison. In our design, we select two cache sizes—256 KB and 1 MB—offering a trade-off between area consumption and performance. Section~\ref{subsec:cache_choice} presents the analysis and experimental results supporting our choice of cache sizes. We investigate four memory channel configurations, varying the number of DIMMs while keeping the number of ranks per DIMM constant. As shown in Figure~\ref{fig:speedup}, we observe that latency is influenced by the OTE parameter sets, where bigger $k$ affects the cache hit rate. All parameter sets perform best with 16 active ranks due to high parallelism.


When the cache size is set to 256KB, the total area consumption is $1.482 \text{mm}^2$. Across different parameter sets, the latency ranges from 16.3 ms to 174.4 ms. Compared to the CPU baseline, we achieve a performance improvement ranging from $3.66\times$ to $39.26\times$.
When the cache size is increased to 1MB, the total area consumption increases to $2.995 \text{mm}^2$. In this case, latency ranges from 2.7 ms to 127.0 ms across different parameter sets. We achieve a performance improvement ranging from $5.03\times$ to $237.04\times$. The best performance improvement is observed with the parameter set that has a single execution output size of $2^{20}$.

We implement the PCG-style OTE protocol on the GPU for a comprehensive comparison. By leveraging GPU acceleration,the OTE protocol delivers a $5.88\times$ improvement in throughput. The latency breakdown shows that SPCOT and LPN account for $44.1\%$ and $50.2\%$ of the total latency, respectively. The observed differences in the breakdown compared to the CPU results, as shown in Figure~\ref{fig:intro}(b), are primarily due to the significantly larger L1 and L2 caches in the A6000 GPU, which provide higher bandwidth for LPN operations. Compared to the GPU implementation, as shown in Figure~\ref{fig:speedup}, Ironman achieves a $40.31\times$ reduction in latency, along with an $84.5\times$ reduction in power consumption.


\subsection{SPCOT Operation Optimization}\label{subsec:spcot}

\begin{figure}[!tb]
    \centering
    \includegraphics[width=1\linewidth]{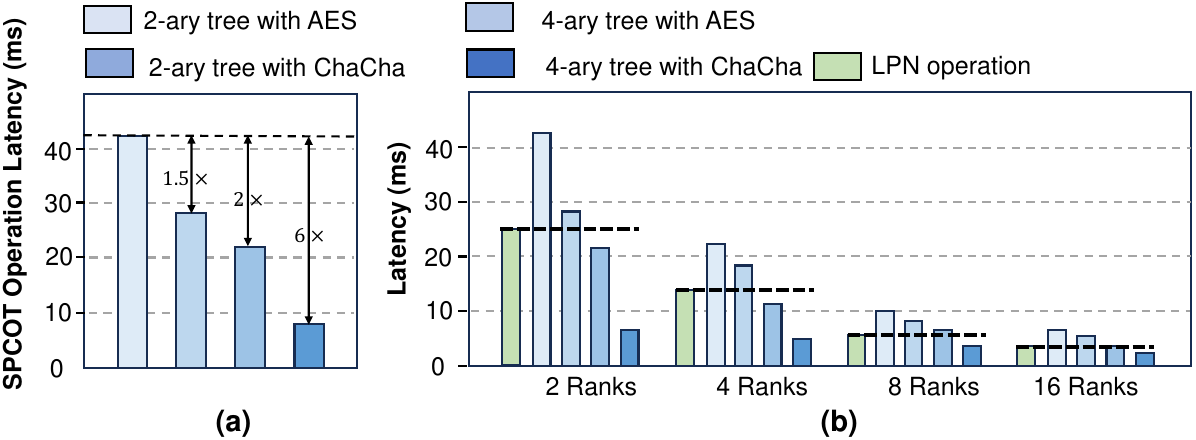}
    \caption{(a) Ablation study of m-ary tree expansion with ChaCha and (b) Latency comparison of SPCOT operation and LPN operation.
    }
    \label{fig:spcot_exp_result}
    \Description{SPCOT Optimization Experiment}
\end{figure}

To demonstrate that the combination of m-ary tree expansion and the selected ChaCha-based PRG achieves significantly better performance, we conduct an ablation study. As shown in Figure~\ref{fig:spcot_exp_result}(a), adopting 4-ary tree expansion with AES or 2-ary tree expansion with ChaCha results in modest performance improvements of $1.5\times$ and $2\times$, respectively. We can further achieve a $6\times$ performance improvement by combining both optimizations. These results highlight the effectiveness of our HW/SW co-design approach without additional hardware cost. 


To demonstrate that accelerating SPCOT improves overall OTE latency, we compare the latency of SPCOT and LPN with varying active ranks. As shown in Figure~\ref{fig:spcot_exp_result}(b), 2-ary and 4-ary tree expansions with AES exceed LPN latency across all channel configurations, dominating the OTE latency. Similarly, 2-ary tree expansion with ChaCha shows the same issue with more ranks. However, 4-ary tree expansion with ChaCha consistently achieves the best performance across different rank configurations, with its latency remaining lower than that of the LPN operation. This result further validates the effectiveness of 4-ary tree expansion with ChaCha in optimizing the overall algorithm.

\subsection{LPN Operation Optimization}
\label{subsec:cache_choice}

\begin{figure*}[ht]
    \centering
    \includegraphics[width=0.90\linewidth]{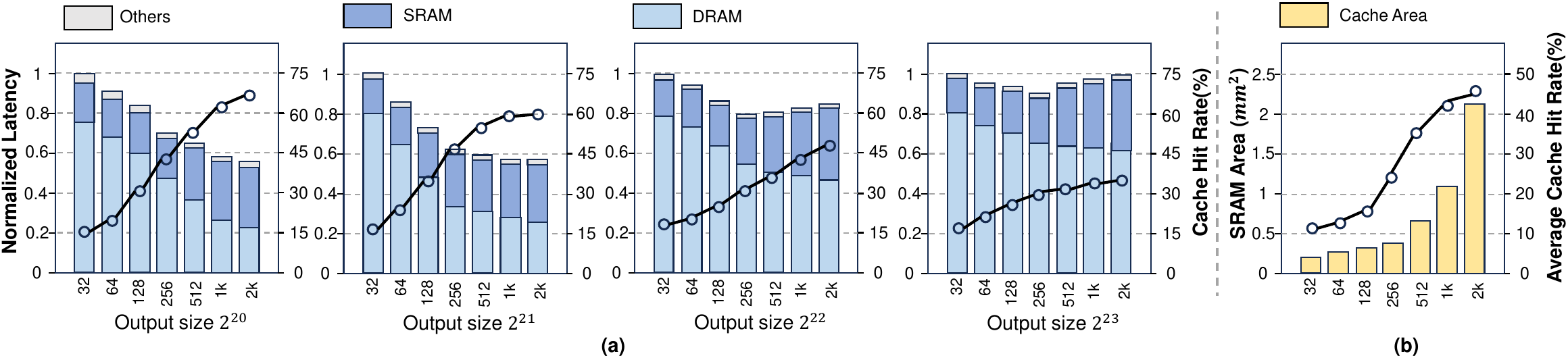}
    \vspace{-3pt}
    \caption{(a) Normalized latency and cache hit rate across different parameter sets, with latency normalized to the 32KB cache size for each parameter set. (b) The average cache hit rate of different parameter configurations for various cache sizes. 
    }
    \label{fig:cache_size}
    \Description{LPN Optimization Experiment}
\end{figure*}

In Figure~\ref{fig:cache_size}(a), we examine the impact of memory-side cache capacity (32KB to 2MB) on normalized latency and cache hit rate across different OT output sizes per protocol execution. Here, each cache line is 64 bytes in size, which matches the DRAM's burst length. 
For smaller parameter configurations ($2^{20}$ or $2^{21}$ OTs per execution), a larger cache provides better performance by increasing the cache hit rate.
On the other hand, for larger sets ($2^{22}$ or $2^{23}$ OTs per execution), increasing the cache size yields diminishing returns in cache hit rate improvement and results in a substantial increase in area consumption. The optimal design point is at a 256KB cache size, beyond which further increases in cache size lead to longer cache access latencies and degrade overall performance.


In Figure~\ref{fig:cache_size}(b), we present the average cache hit rate and the corresponding area for varying cache sizes. When the cache size is increased to 256KB, the cache hit rate improves significantly, with a $1.47\times$ increase compared to the 128KB cache,  with only a slight increase in area.
However, increasing cache size from 1MB to 2MB yields limited improvement in hit rate and a $2.21\times$ area increase. Thus, we choose a 1MB cache for smaller parameter sets and a 256KB cache for larger ones.

\subsection{Operation-wise Performance Profiling}\label{subsec:exp_ppml}

\begin{figure*}[h]
\begin{minipage}{0.52\textwidth}
    \centering
    \includegraphics[width=1.0\textwidth]{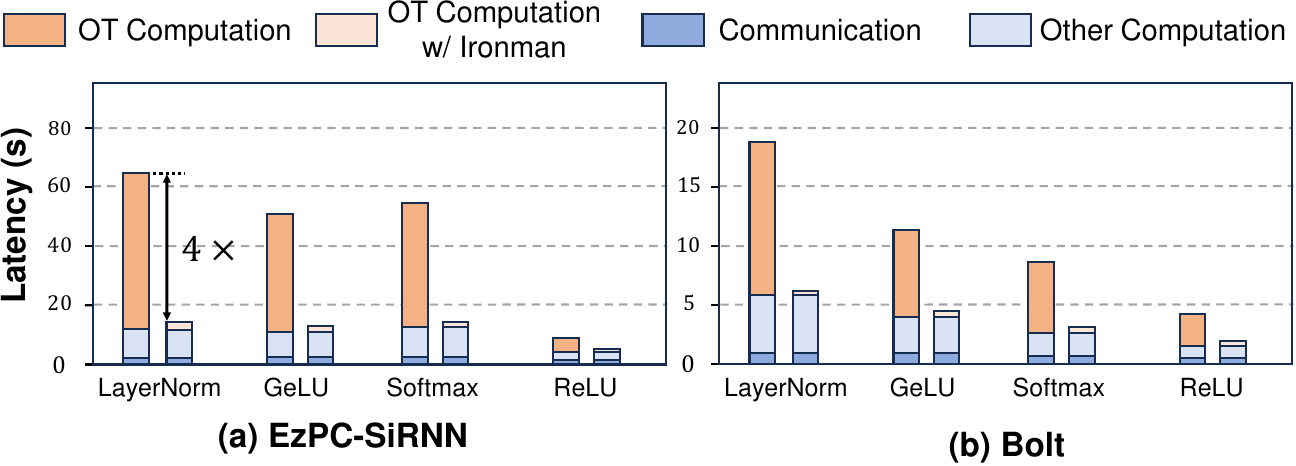} 
\end{minipage}
\hspace{3mm}
\begin{minipage}{0.43\textwidth}
    \centering
    \includegraphics[width=0.80\textwidth]{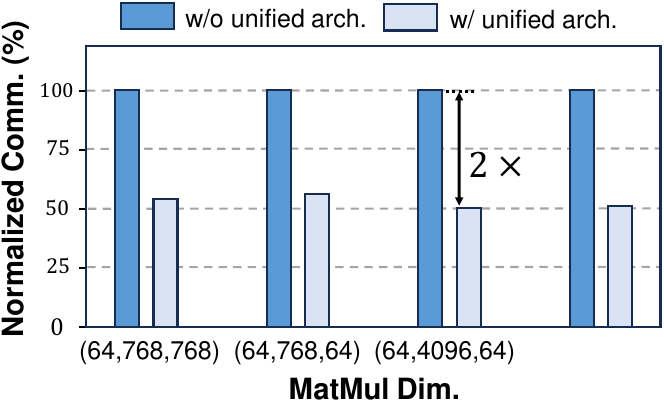} 
\end{minipage}\\
\vspace{-2pt}
\begin{minipage}[t]{.52\textwidth}\centering
    \captionof{figure}{In PPML applications, Ironman can accelerate nonlinear layers based on OT implementations and is applicable to various frameworks including (a) EzPC-SiRNN, (b) Bolt.
    }
    \Description{Non-linear Operation Experiment}
    \label{fig:exp_nonlinear}
\end{minipage}\hspace{3mm}%
\begin{minipage}[t]{.43\textwidth}\centering%
    \captionof{figure}{Communication and latency comparison w/ and w/o unified architecture. The MatMul Dim is represented by the input, hidden, and output dimensions.}
    \label{fig:exp_unified}
    \Description{Unified Unit Experiment}
\end{minipage}
\end{figure*}



We demonstrate Ironman's effectiveness in accelerating PPML applications by optimizing the OT protocol for nonlinear functions. Specifically, we benchmark Ironman on LayerNorm, GeLU, Softmax, and ReLU within two PPML frameworks: EzPC-SiRNN~\cite{rathee2021sirnn,chandran2017ezpc} and Bolt~\cite{pang2024bolt}. As shown in Figure~\ref{fig:exp_nonlinear}, Our experiments show the following key observations: (1) Significant Latency Reduction: Ironman achieves a $3.9\times$ to $4.4\times$ reduction in latency across different nonlinear functions and frameworks. (2) Framework-Agnostic Acceleration: Ironman achieves around a $4\times$ latency reduction across frameworks, primarily  due to OT optimization.

Because we present a specialized hardware design for the host that facilitates seamless switching between sender and receiver roles, we can support the optimizations proposed in Section 4.1 in~\cite{xu2024privquant}, effectively reducing communication costs for linear functions like matrix multiplication (MatMul).
Figure~\ref{fig:exp_unified} shows the results for MatMul across various dimensions. The layer dimensions are derived from the Bert-base model~\cite{devlin2018bert} and LLAMA~\cite{touvron2023llama} with sequence length 32. The results demonstrate that, with the unified architecture, Ironman achieves a $2\times$ reduction in communication, resulting in a $1.4\times$ reduction in latency. 

\subsection{Applications Speedup}
\label{subsec:e2e_app}

\begin{table}[!tb]
    \centering
    \caption{Latency comparison on PPML applications with different bandwidth settings.}
    \label{tab:e2e_result}
    \scalebox{0.72}{
    \vspace{3pt}
    \begin{tabular}{c|c|c|c|c|c|c|c}
    \multicolumn{8}{c}{PPML Application} \\
    \midrule
    \multirow{2}{*}{Frameworks} & \multirow{2}{*}{Models} &  \multicolumn{3}{c|}{400Mbps, 20ms}  & \multicolumn{3}{c}{3Gbps, 0.15ms} \\ 
    & & Base La. & Ours La. & Spd. & Base La. & Ours La. & Spd. \\
    \midrule
    \multirow{6}{*}{CrypTFlow2} & MobileNetV2 & 46.3 & 29.6 & 1.56 & 32.0 & 16.4 & 1.95\\
    & SqueezeNet & 71.0 & 38.8 & 1.83 & 61.8 & 27.7 & 2.23\\
    & ResNet18 & 130.6 & 80.1 & 1.63 & 113.6 & 57.6 & 1.97\\
    & ResNet34 & 287.4 & 168.1 & 1.71 & 217.0 & 100.5 & 2.16\\
    & ResNet50 & 357.4 & 223.5 & 1.60 & 252.4 & 119.7 & 2.11\\
    & DenseNet121 & 629.0 & 411 & 1.53 & 452.5 & 201.3 & 2.25\\
    \midrule
    \multirow{6}{*}{Cheetah} & MobileNetV2 & 31.6 & 22.4 & 1.41 & 12.9 & 5.3 & 2.43\\
     & SqueezeNet & 29.9 & 20.5 & 1.45 & 15.6 & 6.4 & 2.44\\
     & ResNet18 & 39.7 & 27.4 & 1.45 & 21.3 & 9.1 & 2.33\\
     & ResNet34 & 66.1 & 45.4 & 1.47 & 40.7 & 16.3 & 2.49\\
     & ResNet50 & 83.8 & 63.3 & 1.32 & 48.3 & 21.4 & 2.26\\
     & DenseNet121 & 126.9 & 96.5 & 1.33 & 62.1 & 23.3 & 2.67\\
     \midrule
     \multirow{4}{*}{Bolt} & ViT & 1026.8 & 693.8 & 1.48 & 812.2 & 272.6 & 2.98\\
     & BERT-Base & 667.2 & 436.8 & 1.53 & 527.7 & 190.0 & 2.91\\
     & BERT-Large & 1543.2 & 923.9 & 1.67 & 1392.8 & 421.6 & 3.40\\
     & GPT2-Large & 2538.0 & 1555.2 & 1.63 & 2349.4 & 739.4 & 3.18\\
     \bottomrule
     
    \end{tabular}}
    \vspace{-10pt}
\end{table}


We now evaluate \sysname in different hybrid HE/MPC frameworks (including CrypTFlow2~\cite{rathee2020cryptflow2}, Cheetah~\cite{huang2022cheetah}, and Bolt~\cite{pang2024bolt}) for both CNN and Transformer models.
For CNNs, we choose MobileNetV2\cite{sandler2018mobilenetv2}, SqueezeNet\cite{iandola2016squeezenet}, ResNet\cite{he2016deep}, and DenseNet\cite{huang2017densely}, while for Transformer models, we select Vision Transformer (ViT)\cite{dosovitskiy2020image}, BERT\cite{devlin2019bert}, and GPT-2\cite{radford2019language}. We further consider different bandwidth configurations to assess the impact of communication cost and interaction rounds. We set the bandwidth and round-trip latency between cloud instances as (3 GBps, 0.15 ms) and (400 MBps, 20 ms), following \cite{huang2022cheetah}.



Based on the results shown in Table\ref{tab:e2e_result}, three key observations can be made: (1) For the (3 GBps, 0.15 ms) setting, Ironman achieves a latency reduction of $2.11\times$ to $2.67\times$ for CNN-based models and $2.91\times$ to $3.40\times$ for Transformer-based models, with minimal additional hardware resources. These results clearly demonstrate that \sysname can support different frameworks and model sizes, achieving significant performance improvements; (2) Transformer-based models achieve greater performance improvements than CNN-based models. This is because non-linear functions in Transformer-based models, such as Softmax and GeLU, are more complex than those in CNN-based models like ReLU, which consume more OT correlations in the protocol; (3) For the (400 MBps, 20 ms) setting, where bandwidth is lower and round-trip latency is higher, \sysname yields only limited improvements. This is because, after significantly optimizing the OT computation, communication latency between the sender and receiver becomes the new bottleneck for all hybrid PPML frameworks.

\subsection{Overall Area and Power Analysis}\label{subsec:area}

\begin{table}[!tb]
    \centering
    \caption{The design overhead of Ironman-NMP}
    \scalebox{0.95}{
    \label{tab:design_overhead}
    \renewcommand\arraystretch{1.3}
    \resizebox{\linewidth}{!}{
    \begin{tabular}{c|c|c|c|c}
    \hline
    & \multirow{2}{*}{\makecell[c]{ChaCha8 \\ Core}} & \multicolumn{2}{c|}{Ironman-NMP} & \multirow{2}{*}{\makecell[c]{Typical DRAM \\ chip}} \\
    \cline{3-4}
    & & 256KB Cache & 1MB Cache &  \\
    \toprule
    Area ($\text{mm}^2$) & 0.215 & $1.482$ & $2.995$ & $100$ \\
    \midrule
    Power ($W$) & 45.33$mW$ & 1.301 & 1.430 & $10$\\
    \bottomrule
    \end{tabular}
    }
    }
\end{table}

Table~\ref{tab:design_overhead} shows the area and power consumption of Ironman. Under 40nm technology, MetaNMP has a very small area overhead of $1.482\text{mm}^2$ and $2.995\text{mm}^2$ with different sizes of Memory-side Cache, which is much smaller than the $100\text{mm}^2$ area of a typical DRAM chip \cite{dai2022dimmining}. The power consumption of Ironman is $1.301W$ with a 256KB cache and $1.430W$ with a 1MB cache, both of which are significantly smaller than the total power consumption of a Load-Reduced DIMM (LRDIMM) at $10W$.

%% file: docs/conclusion.tex
\section{Conclusion}

In this paper, we introduced \sysname, the first Oblivious Transfer Extension (OTE) accelerator based on Near-Memory Processing (NMP) architecture. 
To accelerate SPCOT operation, we propose a hardware-aware $m$-ary GGM tree expansion along with a combined depth-first and breadth-first traversal strategy. This approach achieves superior performance while maintaining low hardware consumption. To accelerate LPN operation, we utilize the NMP architecture to enhance parallelism. Additionally, we equip each Rank-module with a memory-side cache, integrated with a sorting algorithm, to provide increased bandwidth. To offer greater flexibility in MPC protocol design, we incorporate an XOR tree unit that supports both sender and receiver protocols in a unified manner. With our proposed framework, we achieve a $39.2$–$237.4\times$ improvement in OT throughput and a $2.1$–$3.4\times$ reduction in the latency of privacy-preserving machine learning applications.

\begin{acks}
    This work was supported by the National Natural Science Foundation of China (NSFC) under Grants 62495102, 92464104, and 62125401; the National Key Research and Development Program of China under Grant 2024YFB4505004; the Beijing Municipal Science and Technology Program under Grant Z241100004224015; the 111 Project under Grant B18001; and the Damo Academy Innovative Research Program.
\end{acks}